\documentclass[final,letterpaper,10pt,journal]{IEEEtran} 

\usepackage{amsmath,amssymb,amsfonts,mathtools}
\usepackage{graphicx}
\usepackage{epstopdf}
\usepackage{caption}
\usepackage{subcaption}
\usepackage{cite}
\usepackage{color}

\usepackage{enumerate}
\usepackage[justification=centering]{caption}
\usepackage{makecell}
\usepackage[mathscr]{euscript}
\usepackage{stmaryrd}
\usepackage{multirow}
\usepackage{bm,upgreek}
\usepackage{framed}
\usepackage{diagbox}
\usepackage{algorithm,algorithmic}
\usepackage{textcomp,enumitem}
\usepackage{changepage}

\newcommand{\Cov}{\mathrm{Cov}}

\newcommand{\diff}{\mathrm{d}}

\newtheorem{theorem}{Theorem}
\newtheorem{lemma}{Lemma}
\newtheorem{proposition}{Proposition}
\newtheorem{remark}{Remark}

\newtheorem{assumption}{Assumption}
\newtheorem{corollary}{Corollary}

\renewcommand{\IEEEQED}{\IEEEQEDopen}
\IEEEoverridecommandlockouts
\allowdisplaybreaks[3]
\begin{document}
\title{Continuous-Time Modeling of Destination-Constrained Dynamic Systems via Doob's $h$-Transform
	\thanks{
		The work of Sanfeng Hu was supported by the National Natural Science Foundation of China under Grant 62461058. The work of Liping Guo was supported in part by the National Natural Science Foundation of China under Grant 62603311 and in part by
		the Talent Scientific Fund of Lanzhou University under Grant 561120225. The work of Mengjiao
		Tang was supported by the National Natural Science Foundation
		of China under Grant 62263032 and Grant 12231017. The work of Yao
		Rong was supported by the Youth Project of the “Xingdian” Talent
		Support Plan of Yunnan Province. 
		\emph{(Corresponding author: Liping Guo.)}}
	\thanks{Sanfeng Hu, Mengjiao Tang and Yao Rong are with the Yunnan Key Laboratory of Statistical Modeling and Data Analysis, Yunnan University, Kunming 650504, China
		(e-mail: sanfenghu@outlook.com; qingqing\_tmj@126.com; rongyao0618@outlook.com). Liping Guo is with the School of Mathematics and Statistics, Lanzhou University, Lanzhou 730000, China (e-mail: lipguo@outlook.com). X. Rong Li is with the Department of Electrical Engineering, University of New Orleans, New Orleans, LA
		70148 USA, (e-mail: xli@uno.edu).
		}
}
\author{Sanfeng Hu, Liping Guo, Mengjiao Tang, Yao Rong, and X. Rong Li, \IEEEmembership{Fellow,~IEEE}}

\maketitle
\IEEEpeerreviewmaketitle
\thispagestyle{plain}

\begin{abstract}
	Existing work on modeling destination-constrained dynamic systems has mainly focused on discrete-time formulations with linear destination constraints or Gaussian destination distributions. Since practical systems are more naturally modeled in continuous time and often subject to nonlinear/non-Gaussian constraints/distributions, this paper proposes a unified measure-transformation framework for continuous-time modeling of systems subject to deterministic or distributional constraints on a nonlinear function of the terminal state. The framework starts from an unconstrained It\^{o} stochastic differential equation, typically derived from physical laws to describe the system's forward evolution without destination information, and takes its induced path measure as unconstrained measure. Doob’s $h$-transform is employed to transform this measure into a constrained path measure incorporating destination information through an injected drift. The resulting system is proved to satisfy the constraints while preserving the laws of the unconstrained state process conditioned on the destination information. The transition density is derived, and the path measure of the distributionally constrained system is proved to uniquely minimize Kullback--Leibler divergence from the unconstrained path measure among those satisfying the same distributional constraint. Its injected drift is further shown to converge pointwise to that of the deterministically constrained system as the destination distribution shrinks to the prespecified value. For linear time-varying unconstrained systems, a simplified form of the constrained systems is derived in terms of the conditional expectation of the terminal state, leading to the development of trajectory-generation algorithms. Numerical simulations evaluate the proposed framework under both types of destination constraints.
\end{abstract}

\begin{IEEEkeywords}
	Destination constraints, continuous-time stochastic systems, system modeling, Doob's $h$-transform.
\end{IEEEkeywords}

\section{Introduction}\label{sec:intro}
Mathematical models of dynamic systems arise in numerous fields, such as aerospace and transportation systems \cite{Li1993_ATC, Tomlin1998_ATC}, autonomous robotics \cite{Young1988_Robot}, networked systems \cite{Xue2026, Fioravanti2024}, signal processing \cite{Taghvaei2021}, and cyber-physical security \cite{Guo2026_state}. Physics-based dynamic models characterize forward state transitions but usually do not  incorporate destination information. In many important applications, however, the state process is required to satisfy a prespecified destination constraint at the terminal time \cite{Malyuta2021, Ahmad2017}. Incorporating such information can improve state inference, control and other applications significantly \cite{Li2016compatibility}.

A common approach to incorporating destination information in state estimation is the pseudo measurement method, which conceptually treats the destination as an additional virtual measurement at the terminal time \cite{Tahk1990, Duan2013_Pseudo, Zhou2018}. In this way, classical filters can directly use the virtual measurement to update the state estimates. Another approach is the forward-backward estimate fusion method \cite{CASTANON1985}, which combines a standard forward-filter estimate with a backward prediction initiated from the destination.

As noted in \cite{Rezaie2021_DDT,Xu2022_TSP}, more fundamental for such destination-constrained state inference is to incorporate destination information directly into the state evolution. The resulting constrained model can subsequently be used for inference, control, dynamics analysis, system identification, etc. One such framework is the reciprocal process \cite{Levy1990_Reciprocal, Carravetta2012_Reciprocal, White2011_Reciprocal}. By treating the boundary conditions at both the initial and terminal times as random variables, destination-constrained dynamics can be formalized as Gaussian reciprocal processes \cite{Levy1990_Reciprocal}. Reciprocal models represent two-point boundary constraints but exhibit nearest-neighbor dependencies and dynamically correlated noise, which must be accounted for in sequential work \cite{Levy1990_Reciprocal, Carravetta2012_Reciprocal}. Estimation and smoothing methods have been developed for reciprocal sequences \cite{White2011_Reciprocal, Carravetta2012_Reciprocal} and applied to intent-based anomalous trajectory detection \cite{Fanaswala2013_JSTSP, Fanaswala2014_TSP}. Bridging distribution methods \cite{Ahmad2017, Ahmad2018_CYB} have also been shown to share this reciprocal structure \cite{Rezaie2021_DDT}.

Conditionally Markov sequences impose weaker conditional-independence requirements than reciprocal sequences and therefore accommodate a broader class of endpoint-dependent dynamics \cite{Rezaie2020_TSP, Rezaie2020_TAC_Singular}. A conditionally Markov model specifies a joint density of the initial and terminal states together with a Markov evolution conditional on the terminal state; this factorization retains destination dependence within a one-step forward recursion for modeling, filtering, and prediction \cite{Rezaie2021_DDT}. Related developments cover trajectories with intermediate waypoints \cite{Rezaie2021_Waypoints}, both singular and nonsingular Gaussian sequences \cite{Rezaie2020_TAC_Singular}, and elliptical sequences \cite{Tang2019_ICIF}. Algebraically equivalent dynamic models have also been derived for various conditionally Markov classes \cite{Rezaie2020_TAES_AE, Rezaie2020_TSP}.

An alternative paradigm constructs destination-constrained systems from an unconstrained system. In the linear Gaussian setting, state augmentation and orthogonal decomposition incorporate linear destination constraints into the predictive distribution of the entire state sequence \cite{Xu2022_TSP}. A nearest-neighbor reformulation accommodates both known and uncertain arrival times \cite{Xu2025_TSP}. Under the same special setting, the resulting transition density coincides with that of the pseudo-observation-based bridge distribution model \cite{Xu2025_TSP, Liang2019_SPL}. Subsequent extensions apply this modeling approach to nonlinear unconstrained systems with uncertain arrival times and destinations and to linear unconstrained systems with unknown-but-bounded disturbances in a set-valued framework \cite{Yang2026_TSP, Yang2026_TAC}.

Existing work on destination-constrained dynamic systems predominantly uses discrete-time formulations with linear destination constraints or Gaussian destination distributions. Because physical systems evolve in continuous time, Itô stochastic differential equations (SDEs) provide a more natural and direct description of their underlying dynamics \cite{karatzas1998brownian, sarka2019applied}, and so are advantageous. For example, it avoids introducing time-discretization errors in the model \cite{Arasaratnam2010_TSP}, accommodates irregularly sampled observations \cite{Li2003_TAES}, and enables rigorous system analysis, including those using measure-theoretic tools \cite{Daum1986_TAC, Mitter1996_CSM}. Also, practical applications, such as terminal impact angle control \cite{Ryoo2005_JGCD} and curved manifold navigation \cite{Kingston2018}, may impose nonlinear conditions on the terminal state. These important considerations motivate us to investigate how destination constraints imposed on nonlinear transformations of the terminal state can be incorporated into continuous-time systems. Within  the  framework  of constructing  destination constrained systems from an unconstrained system, we consider two representative types: (a) deterministic constraints, which require the transformed terminal state to lie on a specified nonlinear manifold, and (b) distributional constraints, which require it to follow a prespecified probability distribution.

Extending existing discrete-time approaches to handle these complex destination constraints in the continuous-time domain is challenging. For instance, current methods primarily rely on reformulating step-by-step state transitions, and thus they cannot directly formulate the instantaneous drift required to embed constraints into the continuous evolution described by an SDE.

To address this issue, we shift the perspective from local state transitions to measure transformations on the continuous path space. By treating the path law induced by the unconstrained system as the unconstrained path measure, imposing a destination constraint is equivalent to constructing a constrained path measure under which the state process satisfies the specified destination constraint. The connection between these two measures is established via Doob's $h$-transform, which re-weights the unconstrained path measure based on the likelihood of satisfying the destination constraint. Through Girsanov's theorem, this global path-space re-weighting is translated into an injected drift, yielding an SDE that governs the constrained state process. This formulation directly incorporates general nonlinear destination constraints, both deterministic and distributional, into the continuous-time dynamic system. Our contributions are summarized as follows:

\begin{enumerate}
	\item We establish a unified continuous-time measure-transformation framework for nonlinear systems subject to general nonlinear destination constraints. By applying Doob's $h$-transform, we incorporate deterministic and distributional constraints of the destination into the system evolution through injected drifts and prove that the resulting state processes satisfy the constraints.
	
	\item We derive the transition densities of both constrained systems and establish the information-theoretic optimality of the distributionally constrained system. Both constrained systems preserve the conditional path laws of the unconstrained process. The distributionally constrained path measure is proved to uniquely minimize Kullback–Leibler (KL) divergence from the unconstrained path measure among all path measures satisfying the same distributional constraint. The corresponding injected drift is further shown to converge pointwise to that of the deterministically constrained system as the destination distribution shrinks to the prespecified value.
	
	\item We derive a simplified form of the constrained systems for linear time-varying (LTV) unconstrained systems. This form expresses the injected drifts in terms of the conditional expectation of the terminal state, leading to the development of trajectory-generation algorithms based on Monte Carlo integration and cubature-based linear minimum mean-square error (LMMSE) approximations. We also show that, for the special case of deterministic linear destination constraints, the discretized system coincides with an existing discrete-time destination-constrained system. Finally, we evaluate the proposed framework through numerical simulations under both types of destination constraints.
\end{enumerate}

The paper is organized as follows. Section \ref{sec:preliminaries} reviews the prerequisites on SDEs and Doob's $h$-transform. Section \ref{sec:problem:formulation} formulates the modeling problem for deterministic and distributional destination constraints, and states corresponding assumptions. Section \ref{sec:nonlinear SDE} constructs the two constrained systems and establishes their theoretical properties. Section \ref{sec:linear:system} specializes the framework to LTV unconstrained systems and develops corresponding trajectory-generation algorithms. Section \ref{sec:simulation} evaluates the proposed framework through numerical examples. Finally, Section \ref{sec:conclusion} concludes the paper.

\textit{Notations:} Throughout this paper, $\mathbb{R}^n$ denotes the $n$-dimensional Euclidean space. $\| \cdot \|$ represents the standard Euclidean norm or the induced matrix norm. $A \succ 0$ indicates that $A$ is positive definite. The symbol $\triangleq$ stands for equality by definition. For a scalar function $f$, $\nabla_x f$ denotes its gradient with respect to $x$. For a vector-valued mapping $F=(f_1,\ldots,f_m)^\top$, denote $\nabla_x F \triangleq [\nabla_x f_1,\ldots,\nabla_x f_m]\in\mathbb{R}^{n\times m}$. The symbols $\partial_t f$ and $\partial_{x_i} f$ represent the partial derivatives with respect to time $t$ and the $i$-th coordinate of $x$, respectively. $C^1$ denotes the space of continuously differentiable functions, and $C^{1,2}([0,T] \times \mathbb{R}^n)$ refers to the space of functions that are once continuously differentiable with respect to $t \in [0,T]$ and twice continuously differentiable with respect to $x \in \mathbb{R}^n$. $\mathbb P$ denotes the probability measure and $\mathbb{E}^{\mathbb{P}}[\cdot]$ represents the mathematical expectation taken with respect to $\mathbb{P}$. The symbol $p$ denotes a density under $\mathbb{P}$, whenever it exists. A subscript $Z$ identifies the random vector whose density is denoted by $p_Z$. For $0\le s<t\le T$, define the state transition density $p_{s,t}(\xi | x)\triangleq p_{X_t}(\xi | X_s=x), \ x,\xi\in\mathbb R^n$. When state values carry time subscripts, we write the transition density $p(x_t | x_s)\triangleq p_{X_t}(x_t | X_s=x_s)$ for simplicity; more generally, conditioning on $x_s$ means conditioning on $X_s=x_s$. The same conventions apply to $q, q^*, q^\pi$ under the corresponding transformed measures. The indicator function of an event $A$ is denoted by $\mathbf{1}_A$. $\mathcal{N}(\mu, \Sigma)$ represents a Gaussian distribution with mean vector $\mu$ and covariance matrix $\Sigma$, and $\delta(\cdot)$ is the Dirac delta function. For two probability measures $\mathbb{P}$ and $\mathbb{Q}$ defined on the same measurable space, $\mathbb{Q} \ll \mathbb{P}$ denotes that $\mathbb{Q}$ is absolutely continuous with respect to $\mathbb{P}$, meaning that for any measurable event $A$, $\mathbb{P}(A) = 0$ implies $\mathbb{Q}(A) = 0$. Furthermore, $\mathbb{P}$ and $\mathbb{Q}$ are said to be equivalent measures if they are mutually absolutely continuous (i.e., $\mathbb{Q} \ll \mathbb{P}$ and $\mathbb{P} \ll \mathbb{Q}$), which implies they share the same null events.

\section{Preliminaries}\label{sec:preliminaries}

This section briefly reviews the prerequisites on SDEs and Doob's $h$-transform. 

\subsection{SDE}

Let $(\Omega, \mathcal{F}, \{\mathcal{F}_t\}_{0 \leq t \leq T}, \mathbb{P})$ be a filtered probability space. Consider an $\mathbb{R}^n$-valued stochastic process $\{X_t\}_{0 \leq t \leq T}$ governed by the following nonlinear It\^{o} SDE:
\begin{equation} \label{eq:general_nonlinear_SDE}
	\diff X_t = f(t, X_t) \diff t + g(t, X_t) \diff W_t, \quad X_0 = x_0, \quad t\in [0, T],
\end{equation}
where $W_t$ is a $d$-dimensional $\mathcal{F}_t$-adapted Brownian motion; the drift $f: [0, T] \times \mathbb{R}^n \to \mathbb{R}^n$ and the diffusion $g: [0, T] \times \mathbb{R}^n \to \mathbb{R}^{n \times d}$ are measurable functions; and $T>0$ is given and deterministic. Define the diffusion matrix as $a(t,x) \triangleq g(t,x)g(t,x)^\top$.

To ensure the well-posedness of the process and the analytical properties of its transition density, the following standard assumptions are imposed.

\begin{assumption}[Regularity and Boundedness] \label{assump:regularity}
	The drift $f(t,x)$ and the diffusion $g(t,x)$ satisfy the global Lipschitz condition in $x$, uniformly with respect to $t$. That is, there exists a constant $K_T>0$ such that
	\begin{equation}
		\|f(t, x) - f(t, \xi)\| + \|g(t, x) - g(t, \xi)\|
		\leq K_T \|x - \xi\|
	\end{equation}
	for all $t\in[0,T]$ and $x, \xi \in \mathbb{R}^n$. Furthermore, $f$ and $a$ are globally uniformly bounded and H\"older continuous. Specifically, there exist constants $M_T,L_T>0$ and $\alpha\in(0,1)$ such that
	\begin{equation}
		\|f(t,x)\|+\|a(t,x)\|\leq M_T
	\end{equation}
	and
	\begin{align}
		&\|f(t,x)-f(s, \xi)\|+\|a(t,x) - a(s, \xi)\| \notag\\
		&\leq L_T \left( |t-s|^{\alpha/2} + \|x - \xi\|^\alpha \right)
	\end{align}
	for all $s,t\in[0,T]$ and $x, \xi \in \mathbb{R}^n$.
\end{assumption}

\begin{assumption}[Uniform Ellipticity] \label{assump:ellipticity}
	The diffusion matrix $a(t,x)$ is uniformly positive definite on $[0,T]\times\mathbb{R}^n$. That is, there exists a constant $\lambda_T>0$ such that
	\begin{equation}
		v^\top a(t,x)v\geq\lambda_T\|v\|^2
	\end{equation}
	for all $v\in\mathbb{R}^n$ and $(t,x)\in[0,T]\times\mathbb{R}^n$.
\end{assumption}

\begin{remark}[Analytical Properties] \label{remark:analytical_guarantees}
	Under Assumptions \ref{assump:regularity} and \ref{assump:ellipticity}, the following analytical properties hold:
	\begin{itemize}
		\item The SDE \eqref{eq:general_nonlinear_SDE} admits a unique, non-explosive strong solution with continuous sample paths $\mathbb{P}$-almost surely \cite[Chapter 5, Theorem 2.9]{karatzas1998brownian}. Moreover, the solution is a Markov process \cite[Chapter 5, Theorem 4.20]{karatzas1998brownian}.
		
		\item A transition density $p_{t,T}(\xi | x)$ exists and belongs to $C^{1,2}([0,T)\times\mathbb{R}^n)$ with respect to $(t,x)$ \cite[Sections 1.6 and 1.7]{friedman1964partial}. It satisfies the two-sided Aronson Gaussian bounds \cite[Theorem 1]{aronson1967bounds}: there exist positive constants $K_1,K_2,c_1,c_2$ such that, for all $x, \xi \in \mathbb{R}^n$ and $t\in[0,T)$,
		\begin{align} \label{eq:aronson_bounds}
			\frac{K_1 \exp\left(-c_1\frac{\|x-\xi\|^2}{T-t}\right)}{(T-t)^{n/2}}
			\leq p_{t,T}
			\leq
			\frac{K_2 \exp\left(-c_2\frac{\|x-\xi\|^2}{T-t}\right)}{(T-t)^{n/2}}.
		\end{align}
		In particular, $p_{t,T}(\xi | x)>0$ for all $x,\xi\in\mathbb{R}^n$ and $t\in[0,T)$. Furthermore, the derivatives of $p_{t,T}$ satisfy the Gaussian upper bounds \cite[Sections 1.6 and 1.7]{friedman1964partial}: for any $i,j\in\{1,\ldots,n\}$, there exist constants $C,c>0$ such that
		\begin{align}
			|\partial_{x_i}p_{t,T}(\xi | x)|
			&\leq
			\frac{C \exp\left(-c\frac{\|x-\xi\|^2}{T-t}\right)}{(T-t)^{(n+1)/2}},
			\label{eq:bound_grad_x}\\
			\max\left\{
			|\partial_{x_i x_j}^2p_{t,T}|,\,
			|\partial_t p_{t,T}|
			\right\}
			&\leq
			\frac{C \exp\left(-c\frac{\|x-\xi\|^2}{T-t}\right)}{(T-t)^{(n+2)/2}}.
			\label{eq:bound_partial_t_hess_x}
		\end{align}
	\end{itemize}
\end{remark}

In view of the path continuity established above, we henceforth work on the canonical path space $\Omega=C([0,T],\mathbb{R}^n)$ endowed with the uniform topology, and let $\mathcal{F}$ be its Borel $\sigma$-algebra. Let $\mathbb{P}$ denote the path measure induced by the solution of \eqref{eq:general_nonlinear_SDE}, hereafter called the unconstrained path measure for not incorporating the destination constraint. This measure serves as a basis for the change of measure below. Furthermore, $\{\mathcal{F}_t\}_{0\leq t\leq T}$ denotes the natural filtration of the coordinate process, that is, $\mathcal{F}_t=\sigma(X_s:0\leq s\leq t)$ for $0\leq t\leq T$. This canonical Polish-space setting ensures the existence of the regular conditional probabilities used below.

\subsection{Doob's $h$-Transform}\label{sec:preliminaries:doob}

Doob's $h$-transform characterizes how the governing SDE of a stochastic process changes under an equivalent measure transformation \cite{rogers2000diffusions, sarka2019applied}. When the Radon--Nikodym derivative between the new measure and the original measure is defined by a strictly positive function $h$, this transformation re-weights the probabilities of the sample paths. Through Girsanov's theorem \cite[Theorem 3.5.1]{karatzas1998brownian}, this change of measure introduces an additional drift term, referred to below as the injected drift, into the original SDE, while the diffusion term remains unchanged.

To formulate Doob's $h$-transform, let $h: [0,T] \times \mathbb{R}^n \to \mathbb{R}$ be a function satisfying the following three conditions:
\begin{enumerate}
	\item \emph{Strict positivity:} $h(t, x) > 0$ for all $(t, x) \in [0,T] \times \mathbb{R}^n$.
	\item \emph{Classical regularity:} $h(t, x)$ is of class $C^{1,2}([0,T] \times \mathbb{R}^n)$.
	\item \emph{Martingale property:} The process $M_t \triangleq h(t, X_t)$ is a martingale on $[0, T]$ under $\mathbb{P}$ with respect to $\mathcal{F}_t$.
\end{enumerate}

Since $h(t, X_t)$ is a $\mathbb{P}$-martingale, its finite-variation part in the It\^o decomposition must vanish $\mathbb{P}$-almost surely. Applying It\^o's formula \cite[Theorem 3.3.3]{karatzas1998brownian}, the dynamics of $h(t, X_t)$ reduce to its diffusion component:
\begin{align}\label{eq:diff:h}
	\diff h(t, X_t) = \nabla_x h(t, X_t)^\top g(t, X_t) \diff W_t.	
\end{align}
Define a new probability measure $\mathbb{Q}$ on $\mathcal{F}_T$, equivalent to $\mathbb{P}$, via the Radon--Nikodym derivative. Its density process $Z_t$ restricted to $\mathcal{F}_t$ is given by
\begin{align}
	Z_t := \left.\frac{\diff \mathbb{Q}}{\diff \mathbb{P}}\right|_{\mathcal{F}_t} = \frac{h(t, X_t)}{h(0, x_0)}.
\end{align}
Given that $h(t, X_t)$ is a strictly positive $\mathbb{P}$-martingale, $Z_t$ is a well-defined strictly positive martingale satisfying $\mathbb{E}^{\mathbb{P}}[Z_t] = 1$. It follows from \cite[Proposition 5.21]{legall2016brownian} that $Z_t$ admits the Doléans-Dade exponential representation $Z_t = \mathcal{E}(L)_t$, where the local martingale $L_t$ is
\begin{align}
	L_t &= \int_0^t Z_s^{-1} \diff Z_s \notag \\
	&= \int_0^t h^{-1}(s, X_s) \nabla_x h(s, X_s)^\top g(s, X_s) \diff W_s \notag \\
	&= \int_0^t \nabla_x \ln h(s, X_s)^\top g(s, X_s) \diff W_s,   
\end{align}
and \eqref{eq:diff:h} yields the second equality. Girsanov's theorem guarantees that the process $\widetilde{W}_t$ defined by
\begin{equation}
	\diff \widetilde{W}_t = \diff W_t - g(t, X_t)^\top \nabla_x \ln h(t, X_t) \diff t
\end{equation}
is a standard $d$-dimensional Brownian motion under $\mathbb{Q}$. Substituting $\diff W_t = \diff \widetilde{W}_t + g(t, X_t)^\top \nabla_x \ln h(t, X_t) \diff t$ back into the original SDE \eqref{eq:general_nonlinear_SDE}, the dynamics of the state process under $\mathbb{Q}$ are given by
\begin{align}\label{eq:constrainted_nonlinear_SDE}
	\diff X_t = \left( f(t, X_t) + a(t, X_t) \nabla_x \ln h(t, X_t) \right) \diff t + g(t, X_t) \diff \widetilde{W}_t.
\end{align}
Here, $a(t, X_t) \nabla_x \ln h(t, X_t)$ is the injected drift induced by the measure transformation.

To provide a constructive intuition, the effect of Doob's $h$-transform can be interpreted through the deformation of the transition density. Let $p_{t,t+s}(\xi | x)$ denote the transition density of the process \eqref{eq:general_nonlinear_SDE} under $\mathbb{P}$. The $h$-transform induces the new Markov process \eqref{eq:constrainted_nonlinear_SDE} under $\mathbb{Q}$ with a transition density $q_{t,t+s}$ defined by (see e.g., \cite{rogers2000diffusions, sarka2019applied})
\begin{equation} \label{eq:h_transition_density}
	q_{t,t+s}(\xi | x) = p_{t,t+s}(\xi | x) \frac{h(t+s, \xi)}{h(t, x)}.
\end{equation}
Specifically, it alters the transition density, steering the state process toward states associated with larger values of $h$ via the injected drift induced by Girsanov's theorem. This serves as the intuition for constructing $h$ in Section~\ref{sec:nonlinear SDE}.

\section{Problem Formulation}\label{sec:problem:formulation}

The continuous-time state evolution of a dynamic system is typically modeled by the SDE \eqref{eq:general_nonlinear_SDE}. Such models are usually derived from physical laws (e.g., motion models in target tracking) to characterize the forward state transitions. Throughout this paper, we refer to \eqref{eq:general_nonlinear_SDE} as the unconstrained system. Its solution is called the unconstrained state process, whose path law is the unconstrained path measure $\mathbb{P}$.

However, in many applications, destination information is available in the form of a prespecified terminal requirement at time $T>0$ (e.g., an aircraft landing at an airport). Since the unconstrained system does not incorporate this information, its state process does not necessarily satisfy the destination constraint. Therefore, the modeling problem considered is how to use the unconstrained system and the destination information to construct a destination-constrained system. Depending on whether the task requires strict precision or allows for some flexibility, this paper focuses on two types of destination constraints: deterministic constraints and distributional constraints. Let $H:\mathbb{R}^n\to\mathbb{R}^m$, with $m\leq n$, be a $C^1$ mapping used to specify both types of constraints. Write $Y\triangleq H(X_T)$ for the terminal random vector.

\emph{1) Deterministic Constraint:} The system is required to reach a specific destination set at the terminal time. This is formulated as
\begin{equation}\label{eq:destination const}
	X_T \in \mathcal{M}_b \triangleq \{\xi \in \mathbb{R}^n \mid H(\xi) = b\},
\end{equation}
where $b \in \mathbb{R}^m$ and $\mathcal{M}_b \neq \emptyset$. Here, ``deterministic'' refers to the fixed value $Y=b$; the terminal state $X_T$ in $H(X_T) = b$ may remain random within $\mathcal{M}_b$.

\emph{2) Distributional Constraint:} $Y$ is required to follow a prespecified destination distribution with density $\pi$:
\begin{equation}\label{eq:distributional_const}
	Y \sim \pi.
\end{equation}
The positive set of $\pi$, $\mathcal{Y}^+ \triangleq \{y\in\mathbb{R}^m\mid\pi(y)>0\}$, is assumed to contain only attainable values of $H$, i.e., $\mathcal{Y}^+\subseteq H(\mathbb{R}^n)$.

Equations \eqref{eq:destination const} and \eqref{eq:distributional_const} provide a general description of destination constraints through the mapping $H$. When $H$ is invertible, \eqref{eq:destination const} reduces to prescribing the terminal state $X_T=H^{-1}(b)$, and \eqref{eq:distributional_const} reduces to prescribing the distribution of $X_T$.

Accordingly, the objective of this paper is to construct destination-constrained systems satisfying the following requirements:
\begin{enumerate}
	\item The state process satisfies the destination constraint, either almost surely for the deterministic constraint or in distribution for the distributional constraint.
	\item The constrained system preserves the laws of the unconstrained state process conditioned on $Y$, for terminal values admitted by the destination constraint.
\end{enumerate}

\begin{remark}[Interpretation of the Two Modeling Objectives]
	Satisfying the destination constraint alone does not determine the path law of the state process. As discussed in \cite{Li2016compatibility}, a constrained system should preserve the state evolution consistent with both the unconstrained
	system and the constraint. The two modeling objectives follow this idea. The first specifies the terminal value or its distribution. The second preserves the conditional path laws of the unconstrained system for terminal values allowed by
	the constraint. Together, they determine the constrained path law using both the unconstrained system and the destination information.
\end{remark}

To establish these constructions, we impose the following regularity and integrability conditions on the destination constraints.

For the deterministic constraint \eqref{eq:destination const}, we impose the following
conditions to ensure that the destination set forms a non-empty $C^1$ manifold and satisfies the integrability requirements used below:

\begin{assumption}[Constraint Regularity] \label{assump:constraint}
	The gradient matrix of $H$ has full column rank everywhere on $\mathcal{M}_b$; equivalently,
	\begin{equation}
		\det\left(\nabla_x H(\xi)^\top \nabla_x H(\xi)\right) > 0, \quad \forall \xi \in \mathcal{M}_b.
	\end{equation}
	Additionally, there exist constants $K_\rho > 0$ and $q \geq 0$ such that the induced weight function $\rho(\xi) \triangleq \left[\det\left(\nabla_x H(\xi)^\top \nabla_x H(\xi)\right)\right]^{-1/2}$ satisfies the following polynomial growth condition:
	\begin{equation}
		\rho(\xi) \leq K_\rho (1 + \|\xi\|^q), \quad \forall \xi \in \mathcal{M}_b.
	\end{equation}
\end{assumption}

\begin{remark}[Mildness of Constraint Regularity]
	\label{rem:mildness}
	The full-column-rank condition ensures that $\mathcal M_b$ is an
	$(n-m)$-dimensional $C^1$ submanifold
	\cite{Tu2011_Manifolds}. The polynomial growth condition, together with Assumption \ref{assump:exp_integrability}, is used to establish the finiteness and classical regularity of the transformation function $h^*$ constructed below.
	If $\mathcal M_b$ is compact, the
	continuity of $\rho$ implies that it is bounded, and hence the
	growth condition holds with $q=0$. The condition also holds for
	linear or affine constraints, for which $\rho$ is constant.
\end{remark}

\begin{assumption}[Moderate Growth of the Manifold] \label{assump:exp_integrability}
	For any constant $\alpha > 0$, the manifold $\mathcal{M}_b$ satisfies the following Gaussian integrability condition:
	\begin{equation}\label{eq:Gau:inegrability}
		\int_{\mathcal{M}_b} \exp \left( - \alpha \|\xi\|^2 \right) \diff S(\xi) < \infty,
	\end{equation}
	where $\diff S(\xi)$ denotes the $(n-m)$-dimensional Hausdorff measure on $\mathcal{M}_b$.
\end{assumption}

\begin{remark}[Generality of the Growth Condition]
	Assumption \ref{assump:exp_integrability} holds when the surface measure of $\mathcal{M}_b$ grows at most polynomially. Specifically, let $B_r = \{\xi \in \mathbb{R}^n \mid \|\xi\| \le r\}$ denote a closed ball of radius $r$. Suppose that there exist constants $C > 0$ and $q \ge 0$ such that $\int_{\mathcal{M}_b \cap B_r} \diff S(\xi) \le C(1 + r^q)$ for all $r > 0$. Since Gaussian decay dominates polynomial growth, this condition implies \eqref{eq:Gau:inegrability}. It is satisfied, in particular, by compact manifolds and linear or affine hyperplanes.
\end{remark}

For the distributional constraint \eqref{eq:distributional_const}, the following assumption ensures the regularity of the level sets and the integrability required to construct the transformation function.

\begin{assumption}[Distributional-Constraint Regularity and Integrability] \label{assump:target_dist}
	Let $\pi$ be a prespecified density, and define $\mathcal{Y}^+ \triangleq \{y \in \mathbb{R}^m \mid \pi(y)>0\}$ and $\mathcal{M}_y \triangleq \{\xi \in \mathbb{R}^n \mid H(\xi)=y\}$. For every $y\in\mathcal{Y}^+$, the gradient matrix of $H$ has full column rank everywhere on $\mathcal{M}_y$. Define
	\begin{equation}
		\rho_y(\xi)
		\triangleq
		\left[\det\left(\nabla_x H(\xi)^\top\nabla_x H(\xi)\right)\right]^{-1/2},
		\qquad \xi\in\mathcal{M}_y.
	\end{equation}
	There exist constants $K_\rho>0$ and $q\geq0$, independent of $y$, such that
	\begin{equation}
		\rho_y(\xi)\leq K_\rho\left(1+\|\xi\|^q\right),
		\qquad y\in\mathcal{Y}^+,\quad \xi\in\mathcal{M}_y.
	\end{equation}
	Furthermore, for every $\alpha>0$,
	\begin{equation}
		\sup_{y\in\mathcal{Y}^+}
		\int_{\mathcal{M}_y}
		\exp\left(-\alpha\|\xi\|^2\right)\diff S(\xi)
		<\infty,
	\end{equation}
	and
	\begin{align}
		&\int_{\mathcal{Y}^+}
		\frac{\pi(y) 
			\int_{\mathcal{M}_y}
			\exp\left(-\alpha\|\xi\|^2\right)
			\rho_y(\xi)\diff S(\xi)
			}{p_Y(y | x_0)} \diff y
		<\infty.
	\end{align}
	Since $\mathbb{P}(X_0=x_0)=1$, $p_Y(y | x_0)=p_Y(y)$, the density of $Y$ under $\mathbb{P}$, we use $p_Y(y)$ hereafter for $p_Y(y | x_0)$.
\end{assumption}

\begin{remark}[Typical Cases Covered by Assumption
	\ref{assump:target_dist}]
	Assumption \ref{assump:target_dist} is satisfied by linear constraints with compactly supported destination densities. For smooth nonlinear constraints, it is also satisfied when $\pi$ has compact support, the corresponding manifolds are contained in a common compact set, and the gradient matrix has full column rank on these manifolds, provided that $p_Y$ is continuous and positive on the support of $\pi$. Destination densities with unbounded support, including Gaussian densities and Gaussian mixtures, are also allowed whenever the weighted integrability condition holds.
\end{remark}

\section{Destination-Constrained Nonlinear Systems}\label{sec:nonlinear SDE}

In this section, we use Doob's $h$-transform to construct destination-constrained systems satisfying the two requirements formulated in Section \ref{sec:problem:formulation}.

\subsection{Deterministic Destination Constraints Case}

As introduced in Section \ref{sec:preliminaries}, Doob's $h$-transform alters the transition density, steering the state process toward states characterized by larger values of $h$. Intuitively, the likelihood of the state process reaching the destination captures this guiding property. We construct $h^*$ by evaluating $p_Y(y | X_t=x)$, the density of $Y$ conditioned on $X_t=x$, at $y=b$:
\begin{equation} \label{eq:general_h_def}
	h^*(t, x) \triangleq p_Y(b  | X_t=x).
\end{equation}

The following three lemmas establish the finiteness, strict positivity, and classical regularity of $h^*$, together with the martingale property of $h^*(t,X_t)$, thereby justifying the application of Doob's $h$-transform. These lemmas are proved in Appendices \ref{app:strict_positivity}, \ref{app:smoothness}, \ref{app:martingale}, respectively.

\begin{lemma}[Finiteness and Strict Positivity of $h^*$] \label{lemma:strict_positivity}
	Under Assumptions \ref{assump:regularity}--\ref{assump:exp_integrability}, $h^*$ can be expressed as
	\begin{equation} \label{eq:h_coarea_integral}
		h^*(t, x) = \int_{\mathcal{M}_b} p_{t,T}(\xi | x) \rho(\xi) \diff S(\xi),
	\end{equation}
	where the weight function $\rho(\xi)$ is defined in Assumption \ref{assump:constraint}. Furthermore, $0<h^*(t,x)<\infty$ for all $(t,x)\in[0,T)\times\mathbb{R}^n$.
\end{lemma}

\begin{lemma}[Classical Regularity of $h^*$] \label{lemma:smoothness}
	Under Assumptions \ref{assump:regularity}--\ref{assump:exp_integrability}, $h^*(t, x) \in C^{1,2}([0, T) \times \mathbb{R}^n)$. Furthermore, the differential operators can be interchanged with the surface integral \eqref{eq:h_coarea_integral}, i.e., for any $\mathcal{D} \in \{\partial_{x_i}, \partial_{x_i, x_j}^2, \partial_t\}$ with $1 \leq i, j \leq n$, the following equality holds:
	\begin{align}
		\mathcal D 	h^*(t, x) = \int_{\mathcal{M}_b} \mathcal D p_{t,T}(\xi | x) \rho(\xi) \diff S(\xi).
	\end{align}
\end{lemma}

\begin{lemma}[Martingale Property of $h^*$] \label{lemma:martingale}
	Let $M_t \triangleq h^*(t,X_t)$. Under Assumptions \ref{assump:regularity}--\ref{assump:exp_integrability}, $\{M_t\}_{0\leq t<T}$ is a strictly positive continuous martingale under $\mathbb{P}$ with respect to $\{\mathcal{F}_t\}_{0\leq t<T}$.
\end{lemma}

These properties lead to the following destination-constrained system.

\begin{theorem}[System with Deterministic Destination Constraint] \label{thm:general_transformation}
	Under Assumptions \ref{assump:regularity}--\ref{assump:exp_integrability}, define the deterministically constrained path measure $\mathbb{Q}^*(\cdot) \triangleq \mathbb{P}(\cdot  |  Y = b)$ on the $\sigma$-algebra $\mathcal{F}_T$. Under $\mathbb{Q}^*$, $\lim_{t \uparrow T} H(X_t) = b$ almost surely, and the state process is governed on $[0,T)$ by
	\begin{align} \label{eq:general_constrained_SDE}
		\diff X_t &= \Big[ f(t, X_t) + g(t, X_t)g(t, X_t)^\top \nabla_x \ln h^*(t, X_t) \Big] \diff t \nonumber \\
		&\quad + g(t, X_t) \diff \widetilde{W}_t,
	\end{align}
	where $\widetilde{W}_t$ is a standard Brownian motion under $\mathbb{Q}^*$.
\end{theorem}

\begin{IEEEproof}
	For any $\mathcal{F}_t$ with $t < T$, we evaluate the measure of an arbitrary event $A \in \mathcal{F}_t$ under $\mathbb{Q}^*$. To derive an expression for the probability of $A$ conditional on $Y=b$, we consider an arbitrary Borel set $B \in \mathcal{B}(\mathbb{R}^m)$. By the tower property of conditional expectation, the joint probability expands as
	\begin{equation}
		\mathbb{P}\big(A \cap \{Y \in B\}\big) = \mathbb{E}^{\mathbb{P}}\Big[ \mathbf{1}_A \, \mathbb{P}\big(Y \in B  |  \mathcal{F}_t\big) \Big].
	\end{equation}
	Expressing both sides via their corresponding probability density functions and applying Fubini's theorem to interchange the expectation and integration yields
	\begin{align}
		\int_B \mathbb{P}\big(A  |  Y = y\big) p_Y(y) \diff y &= \int_B \mathbb{E}^{\mathbb{P}} \Big[ \mathbf{1}_A \, p_Y(y  |  \mathcal{F}_t) \Big] \diff y.
	\end{align}
	Because this integral equality holds for all Borel sets $B$, the integrands must coincide almost everywhere. Evaluating this identity at $y = b$ and dividing by $p_Y(b)$, we obtain
	\begin{equation}
		\mathbb{Q}^*(A) = \mathbb{P}(A  |  Y = b) = \frac{\mathbb{E}^{\mathbb{P}} \big[ \mathbf{1}_A \, p_Y(b  |  \mathcal{F}_t) \big]}{p_Y(b)}.
	\end{equation}
	
	By the Markov property of the unconstrained state process, $p_Y(b  |  \mathcal{F}_t) = p_Y(b  |  X_t) \triangleq h^*(t, X_t)$. Furthermore, it follows from the definition of $h^*$ that $p_Y(b) = h^*(0, X_0)$. Thus,
	\begin{equation}
		\mathbb{Q}^*(A) = \int_A \frac{h^*(t, X_t)}{h^*(0, X_0)} \diff \mathbb{P}.
	\end{equation}
	Hence, the restriction of $\mathbb{Q}^*$ to $\mathcal{F}_t$ has the martingale density $Z_t \triangleq \frac{\diff \mathbb{Q}^*}{\diff \mathbb{P}}\big|_{\mathcal{F}_t} = \frac{h^*(t, X_t)}{h^*(0, X_0)}$. It follows from Lemmas \ref{lemma:strict_positivity}--\ref{lemma:martingale} that $Z_t$ is a strictly positive martingale on $[0, T-\epsilon]$ under $\mathbb{P}$ with $\mathbb{E}^{\mathbb{P}}[Z_t] = 1$. Therefore, applying Doob's $h$-transform on each truncated interval $[0,T-\epsilon]$ shows that, under $\mathbb{Q}^*$, the state process is governed by \eqref{eq:general_constrained_SDE} on $[0,T)$.
	
	Finally, since $\mathbb{Q}^*$ is concentrated on $\{\omega\in\Omega: Y(\omega)=b\}$, the continuity of the canonical paths and of $H$ implies
	\begin{equation}
		\mathbb{Q}^*\left(\lim_{t\uparrow T}H(X_t)=b\right)=1.
	\end{equation}
	This completes the proof.
\end{IEEEproof}

Thus, the deterministic destination information modifies only the drift of the unconstrained system through the injected drift $u_t^*=g(t,X_t)g(t,X_t)^\top\nabla_x\ln h^*(t,X_t)$.

\subsection{Distributional Destination Constraints Case}

While Theorem \ref{thm:general_transformation} establishes the system for the deterministic constraint $Y = b$, we now extend this measure-transformation framework to accommodate the distributional constraint $Y \sim \pi$ formulated in Section \ref{sec:problem:formulation}.

\begin{theorem}[System with Distributional Destination Constraint] \label{thm:dist_constrained}
	Under Assumptions \ref{assump:regularity}, \ref{assump:ellipticity}, and \ref{assump:target_dist}, define the distributionally constrained path measure $\mathbb{Q}^\pi(\cdot) \triangleq \int_{\mathbb{R}^m} \mathbb{P}(\cdot  |  Y=y)\pi(y) \diff y$ on the $\sigma$-algebra $\mathcal{F}_T$. The measure $\mathbb{Q}^\pi$ is absolutely continuous with respect to $\mathbb{P}$ on $\mathcal{F}_T$, with Radon--Nikodym derivative
	\begin{equation}\label{eq:soft_rn_derivative}
		\left.\frac{\diff \mathbb{Q}^\pi}{\diff \mathbb{P}}\right|_{\mathcal{F}_T}
		= \frac{\pi(Y)}{p_Y(Y)}, \qquad \mathbb{P}\text{-a.s.}
	\end{equation}
	Under $\mathbb{Q}^\pi$, $Y=H(X_T)$ has density $\pi$. The distribution of $H(X_t)$ under $\mathbb{Q}^\pi$ converges weakly to the probability distribution with density $\pi$ as $t \uparrow T$, and the state process is governed on $[0,T)$ by
	\begin{align}
		\diff X_t &= \Big[ f(t,X_t) + g(t,X_t)g(t,X_t)^\top \nabla_x \ln h^\pi(t,X_t) \Big] \diff t \notag \\
		& \quad + g(t,X_t) \diff \widetilde{W}_t,
	\end{align}
	where $\widetilde{W}_t$ is a standard Brownian motion under $\mathbb{Q}^\pi$. The transformation function $h^\pi(t,x)$ is given by
	\begin{equation}
		h^\pi(t,x) \triangleq \int_{\mathcal{Y}^+} \frac{\pi(y) p_Y(y  |  X_t=x)}{p_Y(y)} \diff y.
	\end{equation}
\end{theorem}

\begin{IEEEproof}
	For any $A \in \mathcal{F}_T$, we evaluate its probability under $\mathbb{Q}^\pi$. By definition,
	\begin{equation}
		\mathbb{Q}^\pi(A) = \int_{\mathbb{R}^m} \mathbb{P}\big(A  |  Y=y\big) \pi(y) \diff y.
	\end{equation}
	It follows from Assumption \ref{assump:target_dist} and  Lemma \ref{lemma:strict_positivity} that $p_Y (y)>0$ for every $y\in\mathcal{Y}^+$. Thus,
	\begin{align} \label{eq:soft_rn_proof}
		\mathbb{Q}^\pi(A) &= \int_{\mathcal{Y}^+} \mathbb{P}\big(A  |  Y=y\big) \frac{\pi(y)}{p_Y(y)}  p_Y(y) \diff y \notag \\
		&= \mathbb{E}^{\mathbb{P}}\left[ \mathbf{1}_A \frac{\pi(Y)}{p_Y(Y)} \right].
	\end{align}
	Since this identity holds for all $A \in \mathcal{F}_T$, it establishes the absolute continuity $\mathbb{Q}^\pi \ll \mathbb{P}$ and the Radon--Nikodym derivative \eqref{eq:soft_rn_derivative}. By the definition of $\mathbb{Q}^\pi$, $Y$ has density $\pi$ under $\mathbb{Q}^\pi$.
	
	To derive the continuous-time system, we restrict $\mathbb{Q}^\pi$ to the filtration $\mathcal{F}_t$ for any $t < T$. The corresponding martingale density process $M_t^\pi$ is given by
	\begin{equation}
		M_t^\pi \triangleq \mathbb{E}^{\mathbb{P}}\left[ \frac{\diff \mathbb{Q}^\pi}{\diff \mathbb{P}} \,\middle|\, \mathcal{F}_t \right].
	\end{equation}
	By the Markov property of the unconstrained state process, the conditional expectation simplifies to the proposed transformation function $h^\pi(t, X_t)$:
	\begin{align}
		M_t^\pi &= \mathbb{E}^{\mathbb{P}}\left[ \frac{\pi(Y)}{p_Y(Y)} \;\middle|\; X_t \right] \nonumber \\
		&= \int_{\mathcal{Y}^+} \frac{\pi(y)}{p_Y(y)} p_Y(y  |  X_t) \diff y \notag \\
		&= h^\pi(t, X_t).
	\end{align}
	
	By Assumption \ref{assump:target_dist} and the argument used in the proof of Lemma~\ref{lemma:strict_positivity} in Appendix~\ref{app:strict_positivity}, both $p_Y(y)$ and $p_Y(y |  X_t=x)$ are finite and strictly positive for every $y\in\mathcal{Y}^+$. Hence, $h^\pi(t,x)>0$. Consequently, integrating this conditional density against the nonnegative destination density $\pi(y)$ over $\mathcal{Y}^+$ gives $h^\pi(t,x)>0$ for all $(t,x)\in[0,T)\times\mathbb{R}^n$.

	To establish the $C^{1,2}$ regularity of $h^\pi(t, x)$, consider any differential operator $\mathcal{D} \in \{\partial_{x_i}, \partial^2_{x_i, x_j}, \partial_t\}$. Using the coarea formula, $h^\pi(t, x)$ can be rewritten as
	\begin{align} \label{eq:h_pi_coarea_expanded}
		h^\pi(t, x) 
		&= \int_{\mathcal{Y}^+} \frac{\pi(y)}{p_Y(y)} \left( \int_{\mathcal{M}_y} p_{t,T}(\xi | x) \rho_y(\xi) \diff S(\xi) \right) \diff y \notag \\
		&= \int_{\mathbb{R}^n} p_{t,T}(\xi | x) \frac{\pi(H(\xi))}{p_Y(H(\xi))} \diff \xi.
	\end{align}
	Let $f^\pi(t, x, \xi) \triangleq p_{t,T}(\xi | x) \frac{\pi(H(\xi))}{p_Y(H(\xi))}$. Since the operator $\mathcal{D}$ acts solely on the transition density $p_{t,T}$, we use the same argument as in the proof of Lemma~\ref{lemma:smoothness} in Appendix~\ref{app:smoothness}. For any state $(t, x)$ restricted to an arbitrary compact domain $[0, T-\delta] \times \overline{B_R(0)}$, there exist positive constants $C_{\delta, R}$ and $c$ such that the derivative of the integrand $f^\pi$ is bounded by a dominating function $\Phi^\pi_{\mathcal{D}}(\xi)$:
	\begin{equation} \label{eq:dct_bound_soft}
		|\mathcal{D} f^\pi(t, x, \xi)| \leq \underbrace{ C_{\delta, R} \exp\left( - \frac{c \|\xi\|^2}{2T} \right) \frac{\pi(H(\xi))}{p_Y(H(\xi))} }_{\triangleq \Phi^\pi_{\mathcal{D}}(\xi)}.
	\end{equation}
	
	It remains to verify that $\Phi^\pi_{\mathcal{D}}$ is integrable over $\mathbb{R}^n$. Applying the coarea formula gives
	\begin{align}
		\int_{\mathbb{R}^n} \Phi^\pi_{\mathcal{D}}(\xi) \diff \xi 
		&= C_{\delta, R} \int_{\mathcal{Y}^+} \frac{\pi(y)}{p_Y(y)} \notag \\
		&\qquad \quad  \cdot  \int_{\mathcal{M}_y} \exp\left( - \frac{c \|\xi\|^2}{2T} \right) \rho_y(\xi) \diff S(\xi) \diff y.
	\end{align}
	Taking $\alpha=c/(2T)$, the weighted coarea integrability condition in Assumption \ref{assump:target_dist} implies that the right-hand side is finite. Hence, $\Phi^\pi_{\mathcal{D}}$ is integrable over $\mathbb{R}^n$.
	
	Consequently, the dominated convergence theorem justifies differentiation under the integral sign, yielding
\begin{equation}
		\mathcal{D} h^\pi(t, x) = \int_{\mathcal{Y}^+} \frac{\pi(y)}{p_Y(y)} \mathcal{D} p_Y(y  |  X_t=x) \diff y.
	\end{equation}
	Since $p_{t,T}$ and its required derivatives are continuous in $(t,x)$ and locally dominated by the same integrable function, applying the dominated convergence theorem again shows that $h^\pi$ and the resulting derivatives are continuous. Therefore, $h^\pi\in C^{1,2}([0,T)\times\mathbb{R}^n)$.
	Since $h^\pi$ satisfies all the conditions required for Doob's $h$-transform on $[0,T-\epsilon]$ for every $\epsilon\in(0,T)$, applying the transformation on each such truncated interval yields the destination-constrained system on $[0,T)$ under $\mathbb{Q}^\pi$.
	
	To verify the destination constraint, we consider an arbitrary bounded continuous function $\varphi: \mathbb{R}^m \to \mathbb{R}$ with $|\varphi| \leq M$ for some $M$. For any $t < T$, taking the expectation under $\mathbb{Q}^\pi$ yields
	\begin{equation}
		\mathbb{E}^{\mathbb{Q}^\pi} [\varphi(H(X_t))] = \mathbb{E}^{\mathbb{P}} \left[ \varphi(H(X_t)) \frac{\pi(Y)}{p_Y(Y)} \right].
	\end{equation}
	Since the canonical sample space consists of continuous paths, $\lim_{t \uparrow T} \varphi(H(X_t)) = \varphi(Y)$. The random variable inside the expectation is dominated by $M \frac{\diff \mathbb{Q}^\pi}{\diff \mathbb{P}}$ with $\mathbb{E}^{\mathbb P} [M \frac{\diff \mathbb{Q}^\pi}{\diff \mathbb{P}}] = M < \infty$. Consequently, the dominated convergence theorem justifies passing the limit inside the expectation:
	\begin{align}
		\lim_{t \uparrow T} \mathbb{E}^{\mathbb{Q}^\pi} [\varphi(H(X_t))] 
		&= \mathbb{E}^{\mathbb{P}} \left[ \lim_{t \uparrow T} \varphi(H(X_t)) \frac{\pi(Y)}{p_Y(Y)} \right] \nonumber \\
		&= \mathbb{E}^{\mathbb{P}} \left[ \varphi(Y) \frac{\pi(Y)}{p_Y(Y)} \right] \nonumber \\
		&= \int_{\mathbb{R}^m} \varphi(y) \pi(y) \diff y.
	\end{align}
	The distribution of $H(X_t)$ under $\mathbb{Q}^\pi$ converges weakly to the probability distribution with density $\pi$ as $t \uparrow T$, since the above identity holds for all bounded continuous functions $\varphi$.
\end{IEEEproof}

\subsection{Properties of the Constrained Systems}

Theorems \ref{thm:general_transformation} and \ref{thm:dist_constrained} establish that the constructed systems satisfy the prespecified destination constraints, fulfilling the first modeling requirement. Their path-measure definitions also ensure the second modeling requirement: for the deterministic constraint, $\mathbb{Q}^*(\cdot)=\mathbb{P}(\cdot |  Y=b)$; for the distributional constraint, the definition of $\mathbb{Q}^{\pi}$ retains the conditional path law $\mathbb{P}(\cdot |  Y=y)$ of the unconstrained system and integrates it with the destination density $\pi$. Thus, both constructions preserve the laws of the unconstrained state process conditioned on $Y$. The following theorem characterizes the transition densities of the constrained systems.

\begin{theorem}[Transition Densities of the Constrained Systems]\label{thm:transition_unification}
	Let $q^*(x_t |  x_s)$ and $q^\pi(x_t |  x_s)$ denote the state transition densities under $\mathbb{Q}^*$ and $\mathbb{Q}^\pi$, respectively, for $0\le s<t<T$.
	
	1) Under the conditions of Theorem \ref{thm:general_transformation}, for the deterministic constraint $Y = b$, the transition density turns out to be that of the unconstrained state process conditioned on the terminal event:
	\begin{equation}
		q^*(x_t |  x_s) = p(x_t  |  x_s, Y = b).
	\end{equation}
	
	2) Under the conditions of Theorem \ref{thm:dist_constrained}, for the distributional constraint $Y \sim \pi$, the transition density is governed by the law of total probability over the admissible values of the terminal variable $Y$:
	\begin{equation} \label{eq:soft_transition_density}
		q^\pi(x_t |  x_s) = \int_{\mathcal{Y}^+} p(x_t  |  x_s, Y = y) q_Y^\pi(y  |  x_s) \diff y,
	\end{equation}
	where $q_Y^\pi(y |  x_s)$ denotes the density of
	$Y$ conditioned on $X_s=x_s$ under $\mathbb{Q}^\pi$ and is given by
	\begin{equation}\label{eq:q:pi}
		q_Y^\pi(y |  x_s)
		=
		\frac{\pi(y)p_Y(y |  x_s)}
		{p_Y(y)h^\pi(s,x_s)},
		\qquad y\in\mathcal{Y}^+.
	\end{equation}
\end{theorem}

\begin{IEEEproof}
	1) By Doob's $h$-transform, the transition density under $\mathbb{Q}^*$ is $q^*(x_t |  x_s) = p(x_t |  x_s) \frac{h^*(t, x_t)}{h^*(s, x_s)}$. For the unconstrained state process, applying Bayes' rule and the Markov property to the conditional density yields
	\begin{align} \label{eq:proof_bayes_expand}
		p(x_t  |  x_s, Y = b) &= \frac{p(x_t  |  x_s) p_Y(b  |  x_t)}{p_Y(b  |  x_s)} \notag \\
		&= p(x_t |  x_s) \frac{h^*(t, x_t)}{h^*(s, x_s)}. 
	\end{align}
	Comparing the two expressions concludes the first part.
	
	2) Similarly, the transition density under $\mathbb{Q}^\pi$ is $q^\pi(x_t |  x_s) = p(x_t |  x_s) \frac{h^\pi(t, x_t)}{h^\pi(s, x_s)}$. Expanding $h^\pi(t, x_t)$ using its definition gives
	\begin{align} \label{eq:soft_proof_expansion}
		q^\pi(x_t |  x_s) &= \int_{\mathcal{Y}^+} p(x_t |  x_s) \frac{p_Y(y  |  x_t)}{p_Y(y  |  x_s)} \notag \\
		&\qquad \cdot \frac{\pi(y) p_Y(y  |  x_s)}{p_Y(y) h^\pi(s, x_s)} \diff y.
	\end{align}
	Then, \eqref{eq:soft_transition_density} follows from \eqref{eq:q:pi} and \eqref{eq:proof_bayes_expand}.
\end{IEEEproof}

We next establish an additional property of the distributionally constrained construction: its path measure uniquely minimizes the KL divergence from the unconstrained path measure among admissible path measures with the prespecified distribution of $Y$. We also show that its injected drift converges pointwise to that of the deterministically constrained system as the destination density shrinks to $b$, under the conditions stated below.

\begin{theorem}[KL Optimality and Drift Convergence of the Distributionally Constrained System] \label{thm:kl_divergence_unification}
	Let $\mathcal{Q}_\pi \triangleq \{ \mathbb{Q} \ll \mathbb{P}  |  Y \sim \pi \text{ under } \mathbb{Q} \}$ denote the admissible space of probability measures satisfying the distributional destination constraint. The following statements hold:
	
	1) Suppose that the conditions of Theorem \ref{thm:dist_constrained} hold and
	\begin{equation}\label{eq:finite_terminal_KL}
		D_{\mathrm{KL}}\left(
		\pi\parallel p_Y
		\right)<\infty,
	\end{equation}
	then the measure $\mathbb{Q}^\pi$ established in Theorem \ref{thm:dist_constrained} uniquely minimizes $D_{\mathrm{KL}}(\mathbb{Q} \parallel \mathbb{P})$ over all $\mathbb{Q} \in \mathcal{Q}_\pi$.
	
	2) Under Assumptions \ref{assump:regularity}--\ref{assump:exp_integrability}, for the deterministic constraint $Y=b$, let $\{\pi_\epsilon\}_{\epsilon>0}$ be a family of destination probability densities, each satisfying Assumption \ref{assump:target_dist}, such that
	\begin{equation}
		\operatorname{supp}(\pi_\epsilon)\subseteq B_\epsilon(b)
		\triangleq\{y\in\mathbb{R}^m:\|y-b\|\leq\epsilon\}.
	\end{equation}
	For each $\epsilon>0$, define
	\begin{align}
		h^{\epsilon}(t,x)
		&=\int_{B_\epsilon(b)}\pi_\epsilon(y)
		\frac{p_Y(y |  X_t=x)}{p_Y(y)}\diff y,\label{eq:h_pi_epsilon}\\
		u^{\epsilon}(t,x)
		&=g(t,x)g(t,x)^\top\nabla_x\ln h^{\epsilon}(t,x).
	\end{align}
	Suppose that, for every fixed $(t,x)\in[0,T)\times\mathbb{R}^n$, $p_Y(y |  X_t=x)$ and $\nabla_x p_Y(y |  X_t=x)$ are continuous in $y$ at $b$. Then, for every $(t,x)\in[0,T)\times\mathbb{R}^n$,
	\begin{align}
		\lim_{\epsilon\to0}u^{\epsilon}(t,x)=u^*(t,x),
	\end{align}
	where $
	u^*(t,x)\triangleq g(t,x)g(t,x)^\top\nabla_x\ln h^*(t,x)$.
\end{theorem}

\begin{IEEEproof}
	1) By \eqref{eq:soft_rn_derivative} and the fact that $Y \sim \pi$ under $\mathbb{Q}^\pi$,
	\begin{align}\label{eq:dist_path_KL}
		D_{\mathrm{KL}}(\mathbb{Q}^\pi\parallel\mathbb{P})
		=\mathbb{E}^{\mathbb{Q}^\pi}\left[
		\ln\frac{\pi(Y)}{p_Y(Y)}
		\right] 
		=D_{\mathrm{KL}}(
		\pi\parallel p_Y)<\infty.
	\end{align}
	It suffices to show that an arbitrary $\mathbb{Q}\in\mathcal{Q}_\pi$ satisfying $D_{\mathrm{KL}}(\mathbb{Q}\parallel\mathbb{P})<\infty$ obeys
	\begin{equation*}
		D_{\mathrm{KL}}(\mathbb{Q}\parallel\mathbb{P})
		\ge D_{\mathrm{KL}}(\mathbb{Q}^\pi\parallel\mathbb{P}),
	\end{equation*}
	with equality if and only if $\mathbb{Q}=\mathbb{Q}^\pi$. To apply the chain rule for Radon--Nikodym derivatives, we first verify that $\mathbb{Q}\ll\mathbb{Q}^\pi$.
	
	Suppose there exists an event $A \in \mathcal{F}_T$ such that $\mathbb{Q}^\pi(A) = 0$. By the definition of $\mathbb{Q}^\pi$, we have
	\begin{align}
		\mathbb{Q}^\pi(A) = \mathbb{E}^{\mathbb{P}}\left[ \mathbf{1}_A \frac{\pi(Y)}{p_Y(Y)} \right] = 0.	
	\end{align}
	This implies
	\begin{equation}
		\mathbb{P}\bigl(A\cap\{\pi(Y)>0\}\bigr)=0.
	\end{equation}
	Let $N \triangleq \{y \in \mathbb{R}^m \mid \pi(y) = 0\}$. Equivalently, $\mathbb{P}(A \cap \{Y \notin N\}) = 0$. Since $\mathbb{Q} \ll \mathbb{P}$ by definition of $\mathcal{Q}_\pi$, it follows that $\mathbb{Q}(A \cap \{Y \notin N\}) = 0$. 
	
	Furthermore, because the random variable $Y$ follows the distribution $\pi(y)$ under $\mathbb{Q}$, the measure of the pre-image of the null set $N$ is zero:
	\begin{align}
		\mathbb{Q}(Y \in N) = \int_N \pi(y) \diff y = 0.
	\end{align}
	Therefore, $\mathbb{Q}(A) \leq \mathbb{Q}(A \cap \{Y \notin N\}) + \mathbb{Q}(Y \in N) = 0$. This establishes that $\mathbb{Q} \ll \mathbb{Q}^\pi$.
	
	Having verified this absolute continuity, expanding the KL divergence using the chain rule for Radon--Nikodym derivatives gives
	\begin{align}
		D_{\mathrm{KL}}(\mathbb{Q} \parallel \mathbb{P}) &= \mathbb{E}^{\mathbb{Q}}\left[ \ln \left( \frac{\diff\mathbb{Q}}{\diff\mathbb{Q}^\pi} \frac{\diff\mathbb{Q}^\pi}{\diff\mathbb{P}} \right) \right] \notag \\
		&= D_{\mathrm{KL}}(\mathbb{Q} \parallel \mathbb{Q}^\pi)  + \mathbb{E}^{\mathbb{Q}}\left[ \ln \frac{\pi(Y)}{p_Y(Y)} \right].
	\end{align}
	Since $\mathbb{Q} \in \mathcal{Q}_\pi$, the distribution of $Y$ under $\mathbb{Q}$ must be $\pi(y)$. Thus, the second term on the right-hand side is a constant $C_\pi$ independent of the choice of $\mathbb{Q}$:
	\begin{align}
		C_\pi = \int_{\mathcal{Y}^+} \pi(y) \ln \frac{\pi(y)}{p_Y(y)} \diff y = D_{\mathrm{KL}}\big(\pi \parallel p_Y\big)<\infty.
	\end{align}
	Consequently, we obtain the identity $D_{\mathrm{KL}}(\mathbb{Q} \parallel \mathbb{P}) = D_{\mathrm{KL}}(\mathbb{Q} \parallel \mathbb{Q}^\pi) + C_\pi$. Since $D_{\mathrm{KL}}(\mathbb{Q} \parallel \mathbb{Q}^\pi) \geq 0$, with equality if and only if $\mathbb{Q}$ and $\mathbb{Q}^\pi$ coincide on $\mathcal{F}_T$, $\mathbb{Q}^\pi$ is the unique measure that minimizes the KL divergence.

	2) For fixed $(t,x)\in[0,T)\times\mathbb{R}^n$, denote the density ratio in \eqref{eq:h_pi_epsilon} by
	\begin{equation}
		F(t,x,y)\triangleq
		\frac{p_Y(y |  X_t=x)}{p_Y(y)}.
	\end{equation}
	Since $p_Y(b)=h^*(0,x_0)>0$, the assumed continuity implies that the denominator is bounded away from zero on $B_{\epsilon_0}(b)$ for some $\epsilon_0>0$. Hence, $F(t,x,\cdot)$ is continuous at $b$. Moreover, since the denominator is independent of $x$,
	\begin{equation}
		\nabla_xF(t,x,y)
		=
		\frac{\nabla_x p_Y(y |  X_t=x)}
		{p_Y(y)},
	\end{equation}
	$\nabla_xF(t,x,\cdot)$ is also continuous at $b$. For every $0<\epsilon<\epsilon_0$, since $\pi_\epsilon$ is supported on $B_\epsilon(b)$ and integrates to $1$, \eqref{eq:h_pi_epsilon} gives
	\begin{align}
		&\left|h^{\epsilon}(t,x)-F(t,x,b)\right| \notag\\
		&=\left|\int_{B_\epsilon(b)}\pi_\epsilon(y)
		\big( F(t,x,y)-F(t,x,b) \big) \diff y\right| \notag\\
		&\quad\leq\sup_{y\in B_\epsilon(b)}
		\left|F(t,x,y)-F(t,x,b)\right|.
	\end{align}
	By the continuity of $F(t,x,\cdot)$ at $b$, the right-hand side converges to zero as $\epsilon\to0$. Hence,
	\begin{equation}\label{eq:h_pi_epsilon_limit}
		\lim_{\epsilon\to0}h^{\epsilon}(t,x)
		=F(t,x,b)=\frac{h^*(t,x)}{h^*(0,x_0)}.
	\end{equation}
	Moreover, the Gaussian derivative bounds in Remark \ref{remark:analytical_guarantees} and the integrability condition in Assumption \ref{assump:target_dist} provide a locally integrable dominating function. Hence, the dominated convergence theorem justifies differentiation under the integral sign in \eqref{eq:h_pi_epsilon}, yielding
	\begin{equation}
		\nabla_xh^{\epsilon}(t,x)
		=\int_{B_\epsilon(b)}\pi_\epsilon(y)\nabla_xF(t,x,y)\diff y.
	\end{equation}
	Therefore,
	\begin{align}
		&\left\|\nabla_xh^{\epsilon}(t,x)-\nabla_xF(t,x,b)\right\| \notag\\
		&\leq\sup_{y\in B_\epsilon(b)}
		\left\|\nabla_xF(t,x,y)-\nabla_xF(t,x,b)\right\|.
	\end{align}
	The continuity of $\nabla_xF(t,x,\cdot)$ at $b$ then gives
	\begin{equation}\label{eq:grad_h_pi_epsilon_limit}
		\lim_{\epsilon\to0}\nabla_xh^{\epsilon}(t,x)
		=\nabla_xF(t,x,b)
		=\frac{\nabla_xh^*(t,x)}{h^*(0,x_0)}.
	\end{equation}
	Since $h^*(t,x)>0$, combining \eqref{eq:h_pi_epsilon_limit} and \eqref{eq:grad_h_pi_epsilon_limit} yields
	\begin{align}
		\lim_{\epsilon\to0}\nabla_x\ln h^{\epsilon}(t,x)
		&=\frac{\nabla_xh^*(t,x)}{h^*(t,x)}
		=\nabla_x\ln h^*(t,x).
	\end{align}
	Multiplying by $g(t,x)g(t,x)^\top$ gives
	\begin{equation}
		\lim_{\epsilon\to0}u^{\epsilon}(t,x)=u^*(t,x).
	\end{equation}
	This completes the proof.
\end{IEEEproof}

For general nonlinear unconstrained systems, the injected drifts can be difficult to compute. The next section considers the widely used LTV unconstrained systems, for which these drifts can be expressed in terms of the conditional expectation of the terminal state.

\section{Destination-Constrained Linear Systems}\label{sec:linear:system}

This section specializes the proposed framework to LTV unconstrained systems and develops numerical algorithms for generating destination-constrained trajectories.

\subsection{LTV Systems and Regularity Relaxation}

Let the LTV unconstrained system be governed by the following SDE
\begin{equation} \label{eq:linear_SDE}
	\diff X_t = F(t)X_t \diff t + G(t) \diff W_t, \quad X_0 = x_0,
\end{equation}
where $F(t) \in \mathbb{R}^{n \times n}$ and $G(t) \in \mathbb{R}^{n \times d}$ are continuous deterministic matrix-valued functions. 
For \eqref{eq:linear_SDE}, Assumptions \ref{assump:regularity} and \ref{assump:ellipticity} required for the general nonlinear framework can be relaxed. 

Given that $F(t)$ and $G(t)$ are continuous matrix-valued functions on the compact interval $[0, T]$, they are uniformly bounded. Consequently, the linear drift $F(t)x$ and diffusion $G(t)$ satisfy the global Lipschitz and linear growth conditions:
\begin{equation}
	\begin{aligned}
		\|F(t)x - F(t) \xi\| &\leq C_T \|x - \xi \|, \quad \forall x, \xi \in \mathbb R^n, \\
		\|F(t)x\|^2 + \|G(t)\|^2 &\leq \tilde{C}_T(1 + \|x\|^2).
	\end{aligned}
\end{equation}
This guarantees the existence of a unique, non-explosive strong solution to \eqref{eq:linear_SDE} with continuous sample paths $\mathbb{P}$-almost surely \cite[Theorem 5.2.9]{karatzas1998brownian}. Thus, for the LTV system, Assumption \ref{assump:regularity} can be removed.

According to classical linear stochastic system theory \cite[Section 5.6]{karatzas1998brownian}, the solution of the unconstrained system \eqref{eq:linear_SDE} with initial condition $X_t=x$ satisfies
\begin{equation} \label{eq:variation_of_constants}
	X_T = \Phi(T, t)x + \int_t^T \Phi(T, s)G(s) \diff W_s,
\end{equation}
where $\Phi(t, s)$ is the fundamental transition matrix satisfying $\partial_t \Phi(t, s) = F(t)\Phi(t, s)$ with $\Phi(s, s) = I$. Consequently, under $\mathbb{P}$ and conditional on $X_t=x$, the terminal state $X_T$ follows a Gaussian distribution $\mathcal{N}\big(\Phi(T,t)x, \Sigma(t, T)\big)$, with the prediction covariance given by
\begin{equation} \label{eq:linear_covariance_def}
	\Sigma(t, T) \triangleq \int_t^T \Phi(T, s)G(s)G(s)^\top \Phi(T, s)^\top \diff s.
\end{equation}

To ensure non-degenerate Gaussian densities for both terminal
prediction and intermediate state transitions, we impose the
following covariance condition in place of the uniform ellipticity
of Assumption \ref{assump:ellipticity}.

\begin{assumption}[Positive Definiteness of Covariance]
	\label{assump:linear_covariance}
	For any $0\le s<t\le T$, the prediction covariance over $[s,t]$
	is positive definite:
	\begin{align}
		\Sigma(s, t) \triangleq \int_s^t \Phi(t,r)G(r)G(r)^\top \Phi(t,r)^\top \diff r \succ 0.
	\end{align}
\end{assumption}

Under Assumption \ref{assump:linear_covariance}, the law of $X_T$ conditioned on $X_t=x$ admits a strictly positive non-degenerate Gaussian density for every $t<T$. Its  Gaussian form provides derivative bounds required to establish the strict positivity and $C^{1,2}$ regularity of the transformation functions. Therefore, the continuity of $F$ and $G$, together with Assumption \ref{assump:linear_covariance}, replaces Assumptions \ref{assump:regularity} and \ref{assump:ellipticity} for the LTV unconstrained system. Under the corresponding destination-constraint assumptions in Section \ref{sec:nonlinear SDE}, the deterministically/distributionally constrained system construction remains valid. Under this replacement, the conclusions of Theorems \ref{thm:transition_unification} and \ref{thm:kl_divergence_unification} also hold for the LTV unconstrained system whenever their respective conditions are satisfied.

\subsection{Destination-Constrained LTV Systems}

For the LTV unconstrained system, the Gaussian transition density yields a simplified form of the injected drifts in terms of the conditional expectation of the terminal state.

\begin{theorem}[Unified Constrained System via Conditional Expectation] \label{thm:optimal_drift_structure}
	Consider the LTV system \eqref{eq:linear_SDE} subject to either the deterministic constraint $Y = b$ or the distributional constraint $Y \sim \pi$. Suppose that Assumption \ref{assump:linear_covariance} holds. Furthermore, assume that Assumptions \ref{assump:constraint} and \ref{assump:exp_integrability} hold for the deterministic constraint, or Assumption \ref{assump:target_dist} holds for the distributional constraint. Under the constrained path measure, denoted generically by $\mathbb{Q}$, the state process $X_t$ is governed on $[0,T)$ by the following continuous-time system
	\begin{align} \label{eq:optimal_bridge_SDE}
		\diff X_t &= \Big[ F(t)X_t + K(t, T) \big( \hat{x}_T (t, X_t) - \Phi(T, t)X_t \big) \Big] \diff t \notag\\
		&\quad + G(t) \diff \widetilde{W}_t,
	\end{align}
	where $\widetilde{W}_t$ is a standard Brownian motion under $\mathbb{Q}$, and the gain matrix is $K(t, T) \triangleq G(t)G(t)^\top \Phi(T, t)^\top \Sigma(t, T)^{-1}$ for $t<T$. The vector $\hat{x}_T(t, x)$ is the expectation of the terminal state conditioned on the current state $X_t = x$ and the constraint.
	\begin{itemize}
		\item For the deterministic constraint $Y = b$, 
		\begin{equation} \label{eq:manifold_expectation}
			\hat{x}_T (t, x) \triangleq \mathbb{E}^{\mathbb{P}} \big[ X_T  |  X_t = x, Y = b \big].
		\end{equation}
		\item For the distributional constraint $Y \sim \pi$,
		\begin{equation} \label{eq:soft_expectation}
			\begin{aligned}
\hat{x}_T (t,x)&\triangleq\int_{\mathcal{Y}^+}\mathbb{E}^{\mathbb{P}}[X_T |  X_t=x,Y=y] \\
&\qquad\cdot q_Y^\pi(y |  X_t=x)\diff y,
\end{aligned}
		\end{equation}
	where $q_Y^\pi(y |  X_t=x)$ is defined in \eqref{eq:q:pi}.
	\end{itemize}
\end{theorem}

\begin{IEEEproof}
	As introduced above, the transition density $p_{t,T}(\xi | x)$ is Gaussian with mean $\Phi(T,t)x$ and covariance $\Sigma(t,T)$. Thus,
	\begin{align} \label{eq:grad_Gamma}
		&\nabla_x p_{t,T}(\xi | x) \notag\\
		&= p_{t,T}(\xi | x) \Phi(T,t)^\top \Sigma(t, T)^{-1} \big(\xi - \Phi(T,t)x\big).
	\end{align}
	
	\emph{Case 1 (Deterministic Constraint):} Using the Gaussian density and the same dominated-convergence argument as in the proof of Lemma~\ref{lemma:smoothness} in Appendix~\ref{app:smoothness}, we obtain
	\begin{align} \label{eq:grad_h_integral}
		&\nabla_x h^*(t, x) \notag \\
		&= \int_{\mathcal{M}_b} \nabla_x p_{t,T}(\xi | x) \rho(\xi) \diff S(\xi) \notag \\
		&= \Phi(T,t)^\top \Sigma(t, T)^{-1} \notag \\
		&\quad \cdot \int_{\mathcal{M}_b} \big(\xi - \Phi(T,t)x\big) p_{t,T}(\xi | x) \rho(\xi) \diff S(\xi).
	\end{align}
	
	The logarithmic gradient is then obtained:
	\begin{align} \label{eq:log_grad_h_expansion}
		&\nabla_x \ln h^*(t, x) \notag \\
		&= \Phi(T,t)^\top \Sigma(t, T)^{-1} \notag \\
		&\qquad \cdot \left( \frac{\int_{\mathcal{M}_b} \xi \, p_{t,T}(\xi | x) \rho(\xi) \diff S(\xi)}{h^*(t, x)} - \Phi(T,t)x \right).
	\end{align}
	
	The coarea version of the conditional distribution at $b$ is, for every Borel set $A\subseteq\mathbb{R}^n$,
	\begin{align}
		&\mathbb{P}\big(X_T\in A |  X_t=x, Y=b\big) \notag \\
		&=\frac{1}{h^*(t,x)}\int_{A\cap\mathcal{M}_b} p_{t,T}(\xi | x) \rho(\xi)\diff S(\xi).
	\end{align}
	Consequently,
	\begin{align} \label{eq:identified_expectation}
		\hat{x}_T (t,x)
		&=\mathbb{E}^{\mathbb{P}}\big[X_T |  X_t=x, Y=b\big] \notag \\
		&=\frac{1}{h^*(t,x)}\int_{\mathcal{M}_b}\xi\, p_{t,T}(\xi | x) \rho(\xi)\diff S(\xi).
	\end{align}
	
	\emph{Case 2 (Distributional Constraint):} Under $\mathbb{Q}^\pi$, the transformation function is $h^\pi(t,x)$. Using the Gaussian density and the weighted integrability condition in Assumption \ref{assump:target_dist}, the dominated convergence theorem justifies interchanging differentiation and integration, yielding
	\begin{align}
		\nabla_x h^\pi(t, x) &= \int_{\mathbb{R}^n} \nabla_x p_{t,T}(\xi | x) \frac{\pi(H(\xi))}{p_Y(H(\xi))} \diff \xi \notag \\
		&= \Phi(T,t)^\top \Sigma(t, T)^{-1} \notag \\ 
		&\quad \cdot \int_{\mathbb{R}^n} \big(\xi - \Phi(T,t)x\big) r^\pi(t,x,\xi) \diff \xi,
	\end{align}
	where $r^\pi(t,x,\xi) \triangleq p_{t,T}(\xi | x) \frac{\pi(H(\xi))}{p_Y(H(\xi))}$. Dividing by $h^\pi(t,x)$ yields:
	\begin{align}\label{eq:log:pi}
		\nabla_x \ln h^\pi(t, x) &= \Phi(T,t)^\top \Sigma(t, T)^{-1} \notag \\
		&\quad \cdot \left( \int_{\mathbb{R}^n} \xi \, q^\pi_{t,T}(\xi | x) \diff \xi - \Phi(T,t)x \right),
	\end{align}
	where $q^\pi_{t,T}(\xi | x) \triangleq r^\pi(t,x,\xi) / h^\pi(t, x)$. Here $q^\pi_{t,T}(\xi | x)$ is the density of $X_T$ conditioned on $X_t=x$, evaluated at $\xi$, under $\mathbb{Q}^\pi$. By the coarea formula, this integral expands to
	\begin{align}
		&\int_{\mathbb{R}^n} \xi \, q^\pi_{t,T}(\xi | x) \diff \xi \notag \\
		&= \int_{\mathcal{Y}^+}  \int_{\mathcal{M}_y} \xi \, q^\pi_{t,T}(\xi | x) \rho_y(\xi) \diff S(\xi)  \diff y \notag \\
		&= \int_{\mathcal{Y}^+} \frac{\pi(y)}{p_Y(y) h^\pi(t, x)}   \int_{\mathcal{M}_y} \xi \, p_{t,T}(\xi | x) \rho_y(\xi) \diff S(\xi)  \diff y.
	\end{align}
	By the same coarea conditional-distribution argument as in Case 1, the inner surface integral equals $\mathbb{E}^{\mathbb{P}}[X_T  |  X_t=x, Y=y] p_Y(y  |  X_t=x)$. Substituting this expression into the outer integral gives
	\begin{align}
		&\int_{\mathbb{R}^n} \xi \, q^\pi_{t,T}(\xi | x) \diff \xi \notag \\
		&= \int_{\mathcal{Y}^+} \mathbb{E}^{\mathbb{P}} \big[ X_T  |  X_t = x, Y = y \big] \cdot \underbrace{ \frac{\pi(y) p_Y(y  |  X_t=x)}{p_Y(y) h^\pi(t, x)} }_{= q_Y^\pi(y |  X_t=x)} \diff y \notag \\
		&= \hat{x}_T (t, x).
	\end{align}
	Substituting it back into \eqref{eq:log:pi} concludes the proof.
\end{IEEEproof}

For linear destination constraints, the preceding result simplifies as follows.

\begin{corollary}[Constrained Systems with Linear Constraints] \label{cor:linear_deterministic}
	Under the conditions of Theorem \ref{thm:optimal_drift_structure}, let $H(\xi)=A\xi$ where $A \in \mathbb{R}^{m \times n}$ has full row rank, and define the prediction covariance as $\Sigma_A(t, T) \triangleq A \Sigma(t, T) A^\top$. The state process $X_t$ under the constrained path measure is governed on $[0,T)$ by
	\begin{align} \label{eq:bridge_SDE_linear}
		\diff X_t &= \Big[ F(t)X_t + G(t)G(t)^\top \Phi(T, t)^\top A^\top \Sigma_A(t, T)^{-1} \nonumber \\
		&\qquad \cdot \big( \hat{y}(t, X_t) - A\Phi(T, t)X_t \big) \Big] \diff t + G(t) \diff \widetilde{W}_t,
	\end{align}
	where $\hat{y}(t, x)$ is defined as follows:
	\begin{itemize}
		\item For the deterministic constraint $A X_T = b$,
		\begin{equation}
			\hat{y}(t, x) = b, \quad \forall x \in \mathbb{R}^n.
		\end{equation}
		\item For the distributional constraint $A X_T \sim \pi$,
		\begin{equation} \label{eq:y_soft}
			\hat{y}(t, x) = \int_{\mathcal{Y}^+} y q_Y^\pi(y |  X_t=x) \diff y.
		\end{equation}
	\end{itemize}
\end{corollary}

\begin{IEEEproof}
	For $t<T$, conditional on $X_t=x$, the pair $(X_T,AX_T)$ is jointly Gaussian under $\mathbb{P}$. According to the standard property of conditional Gaussian distributions, the conditional mean is
	\begin{align} \label{eq:linear_conditional_mean}
		&\mathbb{E}^{\mathbb{P}}[X_T  |  X_t=x, A X_T = y] \notag \\
		&= \Phi(T, t)x + \Sigma(t, T) A^\top \Sigma_A(t, T)^{-1} \big( y - A \Phi(T, t)x \big).
	\end{align}
	
	Substituting \eqref{eq:linear_conditional_mean} into the definitions of $\hat{x}_T (t, x)$ in \eqref{eq:manifold_expectation} and \eqref{eq:soft_expectation} yields the following expression for both constraint types:
	\begin{equation*}
		\hat{x}_T = \Phi(T, t)x + \Sigma(t, T) A^\top \Sigma_A(t, T)^{-1} \big(\hat{y}(t,x) - A \Phi(T, t)x\big).
	\end{equation*}
	Substituting it into the drift $K(t,T)\big(\hat{x}_T (t,x) - \Phi(T, t)X_t\big)$ leads to \eqref{eq:bridge_SDE_linear}.
\end{IEEEproof}

\subsection{Connection with Discrete-Time Destination-Constrained Systems}
\label{sec:connection_xu2022}

Reference~\cite{Xu2022_TSP} considers discrete-time linear systems with deterministic linear destination constraints. We compare it with the special case $AX_T=b$ of our framework.

For the LTV unconstrained system \eqref{eq:linear_SDE}, let $X_k=X_t|_{t = t_k}$ on the grid $0=t_0<t_1<\cdots<t_N=T$. Define
\begin{align}
 F_k^{\mathrm d}&=\Phi(t_{k+1},t_k),\\
 Q_k^{\mathrm d}&=\int_{t_k}^{t_{k+1}}\Phi(t_{k+1},s)G(s)G(s)^\top
                  \Phi(t_{k+1},s)^\top\diff s,\\
 B_k&=A\Phi(T,t_k),\ S_k=\Sigma_A(t_k,T),\  L_k=Q_k^{\mathrm d}B_{k+1}^\top S_k^{-1},
\end{align}
where $k=0,\ldots,N-1$ and $B_N=A$.

Under $\mathbb P$, the exact discretization of \eqref{eq:linear_SDE} is
\begin{equation}\label{eq:discrete_unconstrained}
 X_{k+1}=F_k^{\mathrm d}X_k+U_k,
\end{equation}
where $\{U_k\}$ is zero-mean white Gaussian noise with covariance $Q_k^{\mathrm d}$.

\begin{proposition}[Relation to the Discrete-Time System]
\label{prop:discrete_dc}
Under the conditions of Corollary~\ref{cor:linear_deterministic}, for the deterministic destination constraint $AX_T=b$, the exact discretization of \eqref{eq:bridge_SDE_linear} on the grid $\{t_k\}$ is
\begin{equation}\label{eq:discrete_dc}
 X_{k+1}=F_k^{\mathrm d}X_k+L_k(b-B_kX_k)+V_k,
\end{equation}
where $\{V_k\}$ is zero-mean white Gaussian noise with covariance
\begin{equation}\label{eq:discrete_dc_cov}
 \overline Q_k=Q_k^{\mathrm d}-L_kS_kL_k^\top.
\end{equation}
For the same initial condition, this system is identical to that in \cite[Eq.~(32)]{Xu2022_TSP} constructed from the unconstrained system \eqref{eq:discrete_unconstrained} with the destination constraint $AX_N=b$.
\end{proposition}

\begin{IEEEproof}
For $AX_T=b$, \eqref{eq:bridge_SDE_linear} is a linear SDE. To apply the standard solution formula for linear SDEs \cite[Section 5.6]{karatzas1998brownian}, define, for $0\le s<t\le T$,
\begin{align}
 \Psi(t,s)&=\Phi(t,s) \notag \\
 &\quad -\Sigma(s,t)\Phi(T,t)^\top A^\top\Sigma_A(s,T)^{-1}A\Phi(T,s).
 \label{eq:constrained_transition_matrix}
\end{align}
Evaluating the solution of \eqref{eq:bridge_SDE_linear} at $t_{k+1}$ gives \eqref{eq:discrete_dc} with
\begin{equation}\label{eq:discrete_dc_noise_integral}
 V_k=\int_{t_k}^{t_{k+1}}\Psi(t_{k+1},s)G(s)\diff\widetilde W_s.
\end{equation}
The integrands are deterministic, and Brownian increments on disjoint intervals are independent. Hence $\{V_k\}$ is zero-mean white Gaussian noise. Under $\mathbb Q^*$, It\^{o}'s isometry and direct integration give the covariance
\begin{align}
 \Cov(V_k)
 &=\int_{t_k}^{t_{k+1}}\Psi(t_{k+1},s)G(s)G(s)^\top \Psi(t_{k+1},s)^\top\diff s\notag\\
 &=Q_k^{\mathrm d}-L_kS_kL_k^\top=\overline Q_k,
\end{align}
which proves \eqref{eq:discrete_dc_cov}.

The matrices in \cite[Eqs.~(23), (26), and (27)]{Xu2022_TSP}, with the index shifted by one, are
\begin{align}
 \Sigma_{k+1,N}&=Q_k^{\mathrm d}\Phi(T,t_{k+1})^\top,\\
 C_{N |  k}^{-1}&=A^\top S_k^{-1}A,\qquad
 \Lambda=A^\top(AA^\top)^{-1}.
\end{align}
Since $A\Lambda=I$ and $B_{k+1}F_k^{\mathrm d}=B_k$, the state equation and noise covariance in \cite[Eqs.~(32) and (44)]{Xu2022_TSP} reduce to \eqref{eq:discrete_dc} and \eqref{eq:discrete_dc_cov}.
\end{IEEEproof}

\subsection{Algorithms for Trajectory Generation}

The implementation of the derived constrained systems requires evaluating the conditional expectation $\hat{x}_T (t, x)$. Formally, applying Bayes' rule, the associated conditional density of the terminal state $X_T$ is proportional to the unconstrained Gaussian prior multiplied by a constraint-induced likelihood:
\begin{equation} \label{eq:bayesian_posterior}
	p_{X_T}(\xi  |  X_t=x, \text{constraint}) \propto \mathcal{N}\big(\xi; \Phi(T, t)x, \Sigma(t, T)\big) \Lambda(\xi),
\end{equation}
where the likelihood function $\Lambda(\xi)$ is defined according to the constraint type:
\begin{equation} \label{eq:likelihood_def}
	\Lambda(\xi) = \begin{cases}
		\delta\big(H(\xi) - b\big), & \text{for } Y = b, \\
		\frac{\pi(H(\xi))}{p_Y(H(\xi))}, & \text{for } Y \sim \pi,
	\end{cases}
\end{equation}
with $\delta(\cdot)$ denoting the Dirac delta function. Due to possible nonlinearity of $H(\cdot)$, the integral defining $\hat{x}_T (t, x)$ is generally not available in closed form. To evaluate this expectation, we present two general numerical frameworks whose implementation can be tailored to the specific destination constraint.

\subsubsection{Monte Carlo Integration \cite{Ristic-Arulampalam-GordonGordon2004, Sarkka2013}}

The conditional expectation can be approximated via importance sampling as $\hat{x}_T (t, x) \approx \sum_{i=1}^{N_p} w^{(i)} X_T^{(i)}$. By drawing $N_p$ particles $X_T^{(i)}$ from the prior $\mathcal{N}\big(\Phi(T, t)x, \Sigma(t, T)\big)$ and setting $Y^{(i)}\triangleq H(X_T^{(i)})$, the normalized importance weights $w^{(i)}$ are assigned according to the likelihood \eqref{eq:likelihood_def}:
\begin{equation} \label{eq:mc_weights}
	w^{(i)} \propto \begin{cases}
		\exp\left( -\frac{\|b - Y^{(i)}\|^2}{2\epsilon} \right), & \text{for } Y = b, \\
		\frac{\pi\big(Y^{(i)}\big)}{p_Y\big(Y^{(i)}\big)}, & \text{for } Y \sim \pi,
	\end{cases}
\end{equation}
where a vanishing variance $\epsilon \to 0$ is introduced to approximate the Dirac delta function for the deterministic constraint. For a distributional constraint, the density $p_Y$ in \eqref{eq:mc_weights} is evaluated using a method tailored to the specific structure of $H$.

\subsubsection{Cubature-Based LMMSE Approximation \cite{Arasaratnam2009Cubature}}

Let $S(t)$ be the Cholesky factor of $\Sigma(t, T)$, i.e., $\Sigma(t, T) = S(t)S(t)^\top$. Using the third-degree spherical-radial cubature rule with points $\gamma_i=\sqrt{n}[I]_i,\
\gamma_{i+n}=-\sqrt{n}[I]_i,\ i=1,\ldots,n$. let
\begin{equation}
	\mathcal{X}^{(i)}(t,x) = \bar{X}(t,x) + S(t) \gamma_i, \  \mathcal{Y}^{(i)}(t,x) = H\big(\mathcal{X}^{(i)}(t,x)\big),
\end{equation}
where $\bar{X}(t,x) \triangleq \Phi(T, t)x$. Then, the required moments at time $t$ are:
\begin{equation}\label{eq:moments}
	\begin{aligned}
\bar{Y}(t,x)&=\frac{1}{2n}\sum_{i=1}^{2n}\mathcal{Y}^{(i)}(t,x), \\
P_{YY}(t,x)&=\frac{1}{2n}\sum_{i=1}^{2n}\mathcal{Y}^{(i)}(t,x)\mathcal{Y}^{(i)}(t,x)^\top -\bar{Y}(t,x)\bar{Y}(t,x)^\top, \\
P_{XY}(t,x)&=\frac{1}{2n}\sum_{i=1}^{2n}\mathcal{X}^{(i)}(t,x)\mathcal{Y}^{(i)}(t,x)^\top -\bar{X}(t,x)\bar{Y}(t,x)^\top.
\end{aligned}	
\end{equation}

Based on the LMMSE criterion, the conditional expectation $\mathbb{E}^{\mathbb{P}}[X_T  |  X_t = x, Y = y]$ is approximated using the moments. Substituting this approximation into \eqref{eq:manifold_expectation} and \eqref{eq:soft_expectation}, the conditional expectation $\hat{x}_T (t, x)$ is evaluated using the following update:
\begin{align} \label{eq:lmmse_update}
\hat{x}_T (t,x)&\approx P_{XY}(t,x)P_{YY}(t,x)^{-1}\big(y^\star(t,x)-\bar{Y}(t,x)\big) \notag \\
&\quad + \bar{X}(t,x),
\end{align}
where 
\begin{itemize}
	\item $y^\star(t,x) = b$, for the deterministic constraint $Y = b$;
	\item $y^\star(t,x) = \int_{\mathcal{Y}^+} y q_Y^\pi(y |  X_t=x) \diff y$, for the distributional constraint $Y \sim \pi$, whose evaluation is problem-dependent and may exploit the specific structures of $H$ and $\pi$.
\end{itemize}

Using either Monte Carlo integration or the cubature-based LMMSE approximation to evaluate $\hat{x}_T (t, x)$, an overall algorithm for destination-constrained trajectory generation is given as Algorithm \ref{alg:trajectory_generation}.

\begin{algorithm}[!t]
	\small
	\caption{Destination-Constrained Trajectory Generation}\label{alg:trajectory_generation}
	\begin{algorithmic}[1]
		\REQUIRE Initial state $X_0$, terminal time $T$, time step $\Delta t$, number of trajectories $M$, Monte Carlo particle number $N_p$ when applicable, nonlinear mapping $H(\cdot)$, prespecified value $b$ or destination density $\pi$, system matrices $F(t)$ and $G(t)$
		\ENSURE $M$ destination-constrained trajectories $\{X_{t_k}^{(m)}: k=0,\ldots,N_t,\; m=1,\ldots,M\}$
		
		\STATE \textbf{Initialize:} $N_t \gets T/\Delta t$
		
		\STATE \textbf{Offline Pre-computation}
		\STATE Compute the discrete-time sequences $\{\Phi(T, t_k)\}_{k=0}^{N_t}$ and $\{\Sigma(t_k, T)\}_{k=0}^{N_t}$ according to Remark \ref{remark:computation_phi_sigma}
		\STATE Cholesky factorization: $\Sigma(t_k, T) = S_k S_k^\top$ for all $t_k < T$
		
		\STATE \textbf{Online Trajectory Generation}
		\FOR{each trajectory $m = 1, 2, \dots, M$}
		\STATE $X_{t_0}^{(m)} \gets X_0$
		\FOR{each time step $k = 0, 1, \dots, N_t-1$}
		\STATE $t_k \gets k\Delta t$
		
		\STATE \textbf{Conditional Expectation:}
		\STATE $x \gets X_{t_k}^{(m)}$
		\IF{using Monte Carlo integration}
		\STATE Evaluate $\hat{x}_T (t_k,x) \gets \sum_{i=1}^{N_p} w^{(i)} X_T^{(i)}$ with \eqref{eq:mc_weights}
		\ELSE
		\STATE Compute $\bar{Y}(t_k,x)$, $P_{YY}(t_k,x)$, and $P_{XY}(t_k,x)$ via \eqref{eq:moments}
		\IF{Deterministic Constraint}
		\STATE $y^\star(t_k,x) \gets b$
		\ELSE 
		\STATE Evaluate $y^\star(t_k,x) \gets \int_{\mathcal{Y}^+} y q_Y^\pi(y |  X_{t_k}=x) \diff y$ using a constraint-specific method
		\ENDIF
		\STATE $\hat{x}_T (t_k,x) \gets P_{XY}(t_k,x) P_{YY}(t_k,x)^{-1} \big( y^\star(t_k,x) - \bar{Y}(t_k,x) \big) + \bar{X}(t_k,x)$
		\ENDIF
		
		\STATE \textbf{Drift Injection:}
		\STATE $u_k^{(m)} \gets G(t_k) G(t_k)^\top \Phi(T, t_k)^\top \Sigma(t_k, T)^{-1} \big(\hat{x}_T (t_k,x) - \Phi(T, t_k)x\big)$
		
		\STATE \textbf{State Integration:}
		\STATE Sample independent Gaussian increment: $Z_k^{(m)} \sim \mathcal{N}(0, I)$
		\STATE $X_{t_{k+1}}^{(m)} \gets X_{t_k}^{(m)} + \big(F(t_k)X_{t_k}^{(m)} + u_k^{(m)}\big)\Delta t + G(t_k)\sqrt{\Delta t}Z_k^{(m)}$
		\ENDFOR
		\ENDFOR
	\end{algorithmic}
\end{algorithm}

\begin{remark}[Computation of $\Phi(T,t)$ and $\Sigma(t,T)$] \label{remark:computation_phi_sigma}
	Algorithm \ref{alg:trajectory_generation} relies on the state transition matrix $\Phi(T,t)$ and the prediction covariance $\Sigma(t,T)$. Depending on the system dynamics, these can be computed as follows:
	\begin{itemize}
		\item \emph{Linear Time-Invariant Systems:} When $F(t) \equiv F$ and $G(t) \equiv G$ are constant matrices, $\Phi(T,t)$ and $\Sigma(t,T)$ can be computed efficiently via the Van Loan's method \cite{van1978computing}. Specifically, by computing the exponential of the $2n \times 2n$ block matrix:
		\begin{equation}
			\exp \left( \begin{bmatrix} -F & GG^\top \\ 0 & F^\top \end{bmatrix} (T-t) \right) = \begin{bmatrix} M_{11} & M_{12} \\ 0 & M_{22} \end{bmatrix},
		\end{equation}
		$\Phi(T,t)$ and $\Sigma(t,T)$ can be represented as
		\begin{equation}
			\Phi(T,t) = M_{22}^\top,\quad \Sigma(t, T) = M_{22}^\top M_{12}.
		\end{equation}
		
		\item \emph{LTV Systems:} For general continuous time-varying matrices $F(t)$ and $G(t)$, both $\Phi(T,t)$ and $\Sigma(t,T)$ are computed by solving a system of matrix ordinary differential equations backward in time from $t=T$. According to the standard adjoint properties of linear systems \cite{rugh1996linear}, the backward evolution is governed by:
		\begin{equation}
			\begin{aligned}
				\frac{\diff}{\diff t} \Phi(T, t) &= -\Phi(T, t)F(t), \quad \Phi(T, T) = I, \\
				\frac{\diff}{\diff t} \Sigma(t, T) &= -\Phi(T, t)G(t)G(t)^\top \Phi(T, t)^\top, \ \Sigma(T, T) = 0.
			\end{aligned}
		\end{equation}
	\end{itemize}
\end{remark}

\section{Numerical Examples} \label{sec:simulation}

Consider a two-dimensional nearly constant velocity motion model as the unconstrained system, governed by the linear SDE $\diff X_t = F X_t \diff t + G \diff W_t$, with the system matrices
\begin{align}
	F = \begin{bmatrix} 	0 & 1 & 0 & 0 \\ 	0 & 0 & 0 & 0 \\ 	0 & 0 & 0 & 1 \\ 	0 & 0 & 0 & 0 \end{bmatrix}, \quad G = \begin{bmatrix} 	0 & 0 \\ 	\sigma & 0 \\ 	0 & 0 \\ 	0 & \sigma \end{bmatrix}.
\end{align}
Here, the state vector is defined as $X_t=[\mathsf{x}_t,\dot{\mathsf{x}}_t,\mathsf{y}_t,\dot{\mathsf{y}}_t]^\top$, where $\mathsf{x}_t$ and $\mathsf{y}_t$ denote the positions along the $\mathsf{x}$- and $\mathsf{y}$-axes, respectively, and $\dot{\mathsf{x}}_t$ and $\dot{\mathsf{y}}_t$ denote the corresponding velocities. In the simulations, the diffusion coefficient is set to $\sigma = 0.2\,\mathrm{m/s^{3/2}}$, the terminal horizon is specified as $T = 50\,\mathrm{s}$, and the discrete time step is $\Delta t = 0.1\,\mathrm{s}$.

The proposed framework is evaluated under three scenarios involving different destination constraints. The first two examples test the framework's ability to handle complex deterministic constraints, specifically focusing on non-convex spatial boundary and disjoint spatial boundaries. The third example evaluates the framework under a distributional constraint, where the terminal position has a prespecified distribution. The cubature-based LMMSE implementation of Algorithm \ref{alg:trajectory_generation} is used for the first two examples, while the Monte Carlo implementation is used for the third example.

\subsection{Downward-Facing Crescent Destination Constraint}

\begin{figure*}[htbp]
	\centering
	\includegraphics[width=0.98\textwidth]{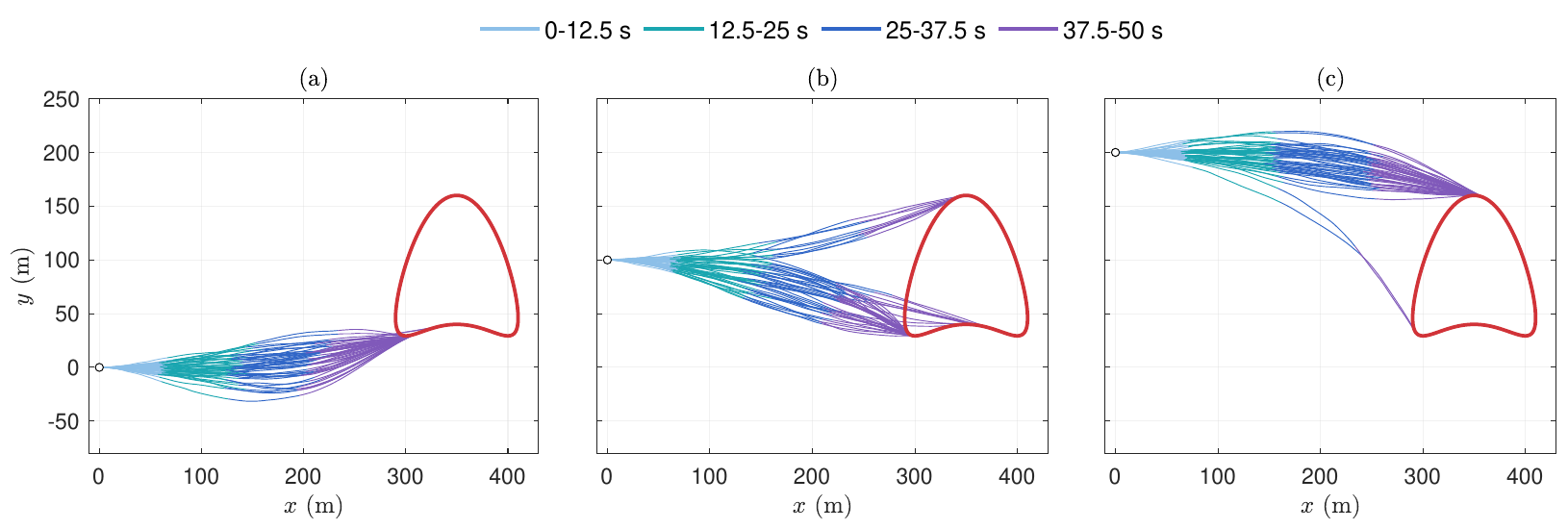}
	\caption{Fifty generated trajectories per initial state, with four colors indicating four consecutive time intervals, as specified in the legend. The red solid curve represents the terminal position constraint $h_1(X_T) = 0$, the boundary of a crescent shape, while a velocity tangency condition $h_2(X_T) = 0$ is simultaneously enforced. The left, middle, and right panels correspond to trajectories initiated from the states $X_0 = [0\,\mathrm{m}, 5\,\mathrm{m/s}, 0\,\mathrm{m}, 0\,\mathrm{m/s}]^\top$, $[0\,\mathrm{m}, 5\,\mathrm{m/s}, 100\,\mathrm{m}, 0\,\mathrm{m/s}]^\top$, and $[0\,\mathrm{m}, 5\,\mathrm{m/s}, 200\,\mathrm{m}, 0\,\mathrm{m/s}]^\top$, respectively.}
	\label{fig:linear_evolution}
\end{figure*}

The system is initialized with a forward horizontal speed of $5\,\mathrm{m/s}$ from various starting altitudes, i.e., $X_0 = [0\,\mathrm{m}, 5\,\mathrm{m/s}, \mathsf{y}_0, 0\,\mathrm{m/s}]^\top$, where $\mathsf{y}_0 \in \{0, 100, 200\}\,\mathrm{m}$. The objective is to steer the stochastic system such that at the terminal time $T = 50\,\mathrm{s}$, the state arrives at a specific non-convex spatial boundary with its velocity tangent to this boundary.

The spatial boundary is designed as a smooth, downward-facing crescent:
\begin{equation}
	h_1(X_T) = (\mathsf{x}_T - c_x)^2 + \big(\mathsf{y}_T - c_y + \kappa(\mathsf{x}_T - c_x)^2\big)^2 - R^2 = 0,
\end{equation}
where $(c_x, c_y)=(350, 100)\,\mathrm{m}$ denotes the geometric center, $R = 60 \,\mathrm{m}$ is the baseline radius, and $\kappa = 1.5\,\mathrm{m}^{-1}$ is the curvature parameter dictating the depth of the concave base. 

To guarantee that the terminal velocity is tangent to this spatial boundary, a tangency condition is enforced. This requires the terminal velocity vector $\mathbf{v}_T = [\dot{\mathsf{x}}_T, \dot{\mathsf{y}}_T]^\top$ to be orthogonal to the curve's normal vector $\mathbf{n}_T = [\nabla_{\mathsf{x}_T} h_1, \nabla_{\mathsf{y}_T} h_1]^\top$. The associated tangency condition is formulated by
\begin{equation}
	h_2(X_T) = \mathbf{n}_T^\top \mathbf{v}_T = 0,
\end{equation}
where
\begin{align}
	\nabla_{\mathsf{x}_T} h_1 &= 2(\mathsf{x}_T - c_x) \Big(1 + 2\kappa \big(\mathsf{y}_T - c_y + \kappa(\mathsf{x}_T - c_x)^2\big) \Big), \notag \\
	\nabla_{\mathsf{y}_T} h_1 &= 2\big(\mathsf{y}_T - c_y + \kappa(\mathsf{x}_T - c_x)^2\big).
\end{align}

The nonlinear destination constraint in the state space is concatenated as $H(X_T) = [h_1(X_T), h_2(X_T)]^\top = 0$, which defines the manifold $\mathcal{M}_{0}$.

As shown in Fig. \ref{fig:linear_evolution}, at the terminal time $T=50\,\mathrm{s}$, the generated trajectories from the three initial states terminate on the non-convex crescent boundary. Their terminal velocities are indeed tangent to the boundary. These numerical results support the simultaneous treatment of the terminal position and velocity constraints.

The trajectory evolution shows how the influence of the injected drift $u_t^*$ changes over time. In the early stages ($t \leq 25\,\mathrm{s}$), the injected drift is weak. As a result, the trajectories disperse due to process noise, behaving similarly to trajectories of the unconstrained state process. As the terminal time approaches, the injected drift increases to overcome this dispersion, steering the trajectories toward the prespecified boundary.

The different trajectory distributions can be explained by the conditional expectation in Theorem \ref{thm:optimal_drift_structure}. The three initial states have the same horizontal components but different initial altitudes, and so their unconstrained terminal means $\Phi(T,0)X_0$ have the same horizontal component and different vertical components. Consequently, the conditional distribution of $X_T$ assigns different weights to points on $\mathcal{M}_{0}$ for the three initial states, and the conditional expectation $\hat{x}_T (t,x)$ directs each group toward the corresponding portions of the crescent.

\subsection{Disjoint Double-Ellipse Constraint}

\begin{figure*}[htbp]
	\centering
	\includegraphics[width=0.98\textwidth]{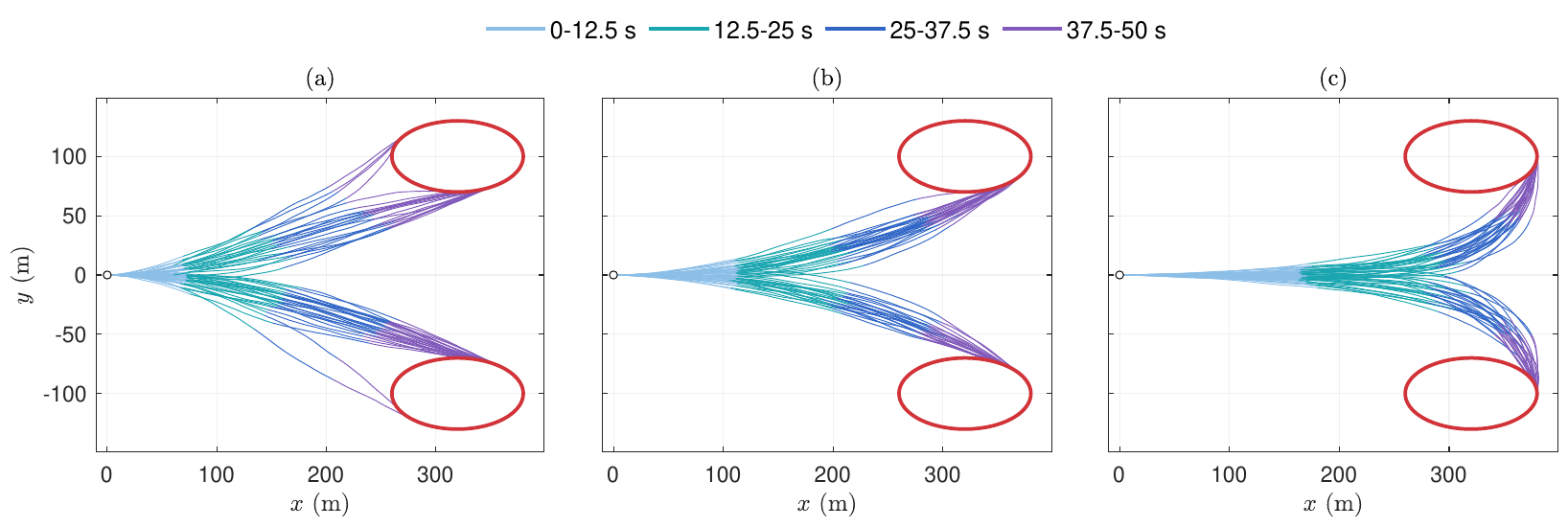}
	\caption{Fifty generated trajectories per initial state, with four colors indicating four consecutive time intervals, as specified in the legend. The red solid curves represent the disjoint terminal position constraint $h_1(X_T) = 0$, one of the ellipses, along with a velocity tangency condition $h_2(X_T) = 0$. The left, middle, and right panels correspond to trajectories initiated from the states $X_0 = [0\,\mathrm{m}, 5\,\mathrm{m/s}, 0\,\mathrm{m}, 0\,\mathrm{m/s}]^\top$, $[0\,\mathrm{m}, 10\,\mathrm{m/s}, 0\,\mathrm{m}, 0\,\mathrm{m/s}]^\top$, and $[0\,\mathrm{m}, 15\,\mathrm{m/s}, 0\,\mathrm{m}, 0\,\mathrm{m/s}]^\top$, respectively.}
	\label{fig:ellipses_evolution}
\end{figure*}

\begin{figure*}[htbp]
	\centering
	\includegraphics[width=0.98\textwidth]{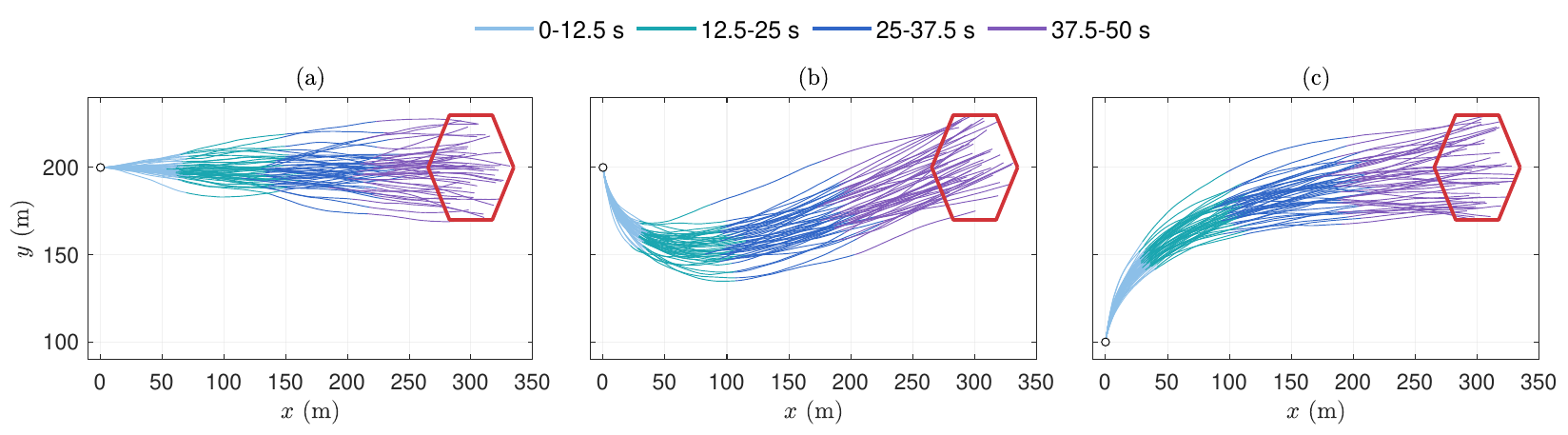}
	\caption{Fifty generated trajectories per initial state, with four colors indicating four consecutive time intervals, as specified in the legend. The red solid hexagon represents the boundary of the region over which the terminal position is uniformly distributed. The left, middle, and right panels correspond to trajectories initiated from the states $X_0 = [0\,\mathrm{m}, 5\,\mathrm{m/s}, 200\,\mathrm{m}, 0\,\mathrm{m/s}]^\top$, $[0\,\mathrm{m}, 1\,\mathrm{m/s}, 200\,\mathrm{m}, -\sqrt{24}\,\mathrm{m/s}]^\top$, and $[0\,\mathrm{m}, 1\,\mathrm{m/s}, 100\,\mathrm{m}, \sqrt{24}\,\mathrm{m/s}]^\top$, respectively.}
	\label{fig:hexagon_evolution}
\end{figure*}

\begin{figure}[htbp]
	\centering
	\includegraphics[width=0.99\columnwidth]{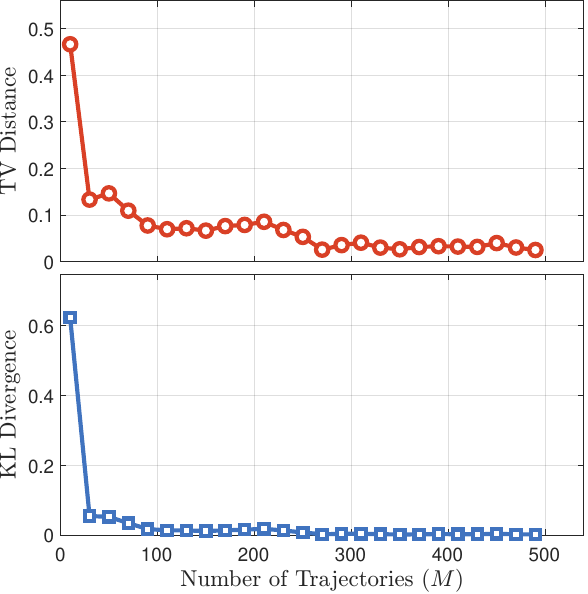}
	\caption{TV distance and KL divergence between the empirical terminal distribution and the prespecified uniform distribution across six equal-area sub-regions, plotted against the number of generated trajectories $M$.}
	\label{fig:Metrics_Convergence}
\end{figure}

In the disjoint spatial boundaries scenario, the system is initialized at the origin $(0, 0)\,\mathrm{m}$ with three distinct initial forward speeds $\dot{\mathsf{x}}_0 \in \{5, 10, 15\}\,\mathrm{m/s}$.

The destination is either of two separated ellipses. This is represented by multiplying two independent elliptical equations: $h_1(X_T) = E_1(X_T) E_2(X_T) = 0$, where $E_i(X_T) = 0$ represents ellipse $i$. The upper and lower ellipses are centered at $(320, 100)\,\mathrm{m}$ and $(320, -100)\,\mathrm{m}$, respectively, with identical semi-axes $a = 60\,\mathrm{m}$ and $b = 30\,\mathrm{m}$. Using the same requirement from the previous scenario, the tangency condition $h_2(X_T) = \mathbf{n}_T^\top \mathbf{v}_T = 0$ is combined with $h_1(X_T)$ to form the final constraint $H(X_T) = 0$.

As shown in Fig. \ref{fig:ellipses_evolution}, at the terminal time $T=50\,\mathrm{s}$, the generated trajectories for each initial condition terminate on either ellipse. Their terminal velocities are indeed tangent to the boundary. As the terminal time approaches, the injected drift separates the trajectories into two groups directed toward the two disjoint boundaries.

This branching behavior can be explained by the conditional expectation in Theorem \ref{thm:optimal_drift_structure}. Initially, the unconstrained state process tends to evolve along the $\mathsf{y}=0\,\mathrm{m}$ axis, which lies between the two boundaries. The conditional expectation then guides each trajectory toward the more appropriate boundary based on its current state. Furthermore, the different initial speeds change the predicted horizontal positions at $T=50\,\mathrm{s}$. As a result, trajectories with a lower initial speed ($\dot{\mathsf{x}}_0=5\,\mathrm{m/s}$) reach the front portions of the ellipses more, while those with a higher speed ($\dot{\mathsf{x}}_0=15\,\mathrm{m/s}$) reach the rear portions more. This result shows how the injected drift accounts for the evolution of the unconstrained state process without requiring a predefined assignment to either boundary.

\subsection{Hexagonal Distributional Constraint}

To evaluate the framework for distributional constraints, the system is initialized from three distinct initial states, as specified in the caption of Fig. \ref{fig:hexagon_evolution}. The destination constraint requires the terminal position to follow a uniform distribution over a regular hexagon centered at $(300, 200)\,\mathrm{m}$ with unconstrained terminal velocity.

Fig. \ref{fig:hexagon_evolution} shows the evolution of the generated trajectories. During the early stage ($t\leq25\,\mathrm{s}$), process noise and the different initial states produce distinct trajectory distributions. As the terminal time approaches, the injected drift drives the trajectories toward the prespecified hexagon. At $T=50\,\mathrm{s}$, the terminal positions are distributed all over the hexagon.

To assess the agreement between the empirical terminal distribution and the prespecified uniform distribution, we partition the hexagon into $K=6$ equal-area triangles. Let $q_i$ denote the empirical probability of the terminal positions in the $i$th triangle for a sample size $M$. Since the destination distribution is uniform, the probability that the terminal position lies in the $i$th triangle is $p_i=1/K$. We then compute the total variation (TV) distance $D_{\mathrm{TV}} = \frac{1}{2} \sum_{i=1}^{K} |q_i-p_i|$ and the KL divergence $D_{\mathrm{KL}} = \sum_{i=1}^{K} q_i \ln(q_i/p_i)$.

As shown in Fig. \ref{fig:Metrics_Convergence}, both metrics decrease toward zero as $M$ increases. This indicates that the empirical terminal probabilities approach the prespecified uniform probabilities over the six-region partition.

\section{Conclusion}\label{sec:conclusion}

This paper has established a measure-transformation framework for the continuous-time modeling of nonlinear systems subject to deterministic/distributional nonlinear destination constraints. Using Doob’s $h$-transform, constrained path measures have been constructed from the unconstrained path measure, and the corresponding constrained systems have been obtained through injected drifts. The resulting state processes have been proved to satisfy the constraints and retain the laws of the unconstrained state process conditioned on $Y$, thereby fulfilling both modeling requirements. Their transition densities have been derived as conditional transition densities of the unconstrained process for deterministic constraints and as integrated densities for distributional constraints. The distributionally constrained path measure has been proved to uniquely minimize KL divergence from the unconstrained path measure among those with the prespecified distribution of $Y$. Its injected drift has also been shown to converge pointwise to that of the deterministically constrained system as the destination distribution shrinks to the prespecified value. For LTV unconstrained systems, a simplified form of the constrained systems has been derived in terms of the conditional expectation of the terminal state, leading to the development of trajectory-generation algorithms based on Monte Carlo integration and cubature-based LMMSE approximations. For deterministic linear destination constraints, the discretized system has also been shown to be identical to an existing discrete-time destination-constrained system. Numerical simulations have demonstrated the effectiveness of the proposed framework under both types of destination constraints.

\appendices

\section{Proof of Lemma~\ref{lemma:strict_positivity}}\label{app:strict_positivity}

	By the coarea formula \cite{evans2015measure}, the coarea version of the conditional density at $b$ is given by
	\begin{equation}
		p_Y(b | X_t=x)
		=
		\int_{\mathcal{M}_b}
		p_{t,T}(\xi | x) \rho(\xi)\diff S(\xi).
	\end{equation}
	
	Fix $(t, x) \in [0, T) \times \mathbb{R}^n$. Consider the integrand $f(\xi) \triangleq p_{t,T}(\xi | x) \rho(\xi)$ on $\mathcal{M}_b$. By Remark \ref{remark:analytical_guarantees}, $p_{t,T}$ is bounded from below by an Aronson Gaussian kernel, ensuring $p_{t,T} > 0$ everywhere on $\mathbb{R}^n$. Furthermore, under Assumption \ref{assump:constraint}, $\rho$ is continuous and strictly positive on $\mathcal{M}_b$. Thus, $f(\xi)$ is continuous and strictly positive for all $\xi \in \mathcal{M}_b$.
	
	Since the manifold $\mathcal{M}_b$ is non-empty, its Hausdorff measure is strictly positive. Integrating the strictly positive function $f(\xi)$ over $\mathcal{M}_b$ yields $h^*(t, x) > 0$.
	
	Moreover, the Aronson upper bound and the polynomial-growth condition on $\rho$ imply that, for some constants $C_{t,x},\alpha_t>0$,
	\begin{equation}
		p_{t,T}(\xi | x)\rho(\xi)
		\leq C_{t,x}(1+\|\xi\|^q)\exp(-\alpha_t\|\xi\|^2).
	\end{equation}
	Choose any $\varepsilon\in(0,\alpha_t)$. Since
	\begin{equation}
		M_{\varepsilon,q}
		\triangleq
		\sup_{\xi\in\mathbb{R}^n}
		(1+\|\xi\|^q)\exp(-\varepsilon\|\xi\|^2)
		<\infty,
	\end{equation}
	we have
	\begin{equation}
		(1+\|\xi\|^q)\exp(-\alpha_t\|\xi\|^2)
		\leq
		M_{\varepsilon,q}
		\exp\bigl(-(\alpha_t-\varepsilon)\|\xi\|^2\bigr).
	\end{equation}
	Assumption \ref{assump:exp_integrability} therefore gives $h^*(t,x)<\infty$. As $(t,x)$ are chosen arbitrarily, $0<h^*(t,x)<\infty$ holds on $[0,T)\times\mathbb{R}^n$.

\section{Proof of Lemma~\ref{lemma:smoothness}}\label{app:smoothness}

	 We evaluate the action of $\mathcal{D}$ on the integrand of \eqref{eq:h_coarea_integral}, denoted as $f(t,x,\xi) \triangleq p_{t,T}(\xi | x) \rho(\xi)$. Since $\mathcal{M}_b$ and $\rho(\xi)$ are independent of $x$ and $t$, the operator acts solely on $p_{t,T}$:
	\begin{equation}
		\mathcal{D} f(t,x,\xi) = \rho(\xi) \mathcal{D} p_{t,T}(\xi | x).
	\end{equation}
	
	According to Remark \ref{remark:analytical_guarantees}, there exist constants $K_{\mathcal{D}} > 0$ and $c > 0$ such that
	\begin{equation} \label{eq:gamma_derivative_bound}
		|\mathcal{D} p_{t,T}(\xi | x) | \leq \frac{K_{\mathcal{D}}}{(T-t)^{\frac{n+r_{\mathcal D}}{2}}} \exp\left( -c \frac{\|x-\xi\|^2}{T-t} \right),
	\end{equation}
	where $r_{\mathcal D}=1$ for $\mathcal D=\partial_{x_i}$, and $r_{\mathcal D}=2$ for $\mathcal D\in\{\partial_{x_i x_j}^2,\partial_t\}$.
	
	To construct a uniform dominating function, we restrict $(t, x)$ to an arbitrary compact domain. Fix a time margin $\delta \in (0, T)$ such that $t \in [0, T-\delta]$, and a spatial radius $R > 0$ such that $x \in \overline{B_R(0)} \triangleq \{x \in \mathbb{R}^n \mid \|x\| \leq R\}$.
	
	By the basic algebraic inequality $\|x-\xi\|^2 \geq \frac{1}{2}\|\xi\|^2 - \|x\|^2$, and substituting $\|x\| \leq R$, we establish an upper bound for the exponent:
	\begin{equation}
		-\frac{c \|x-\xi\|^2}{T-t} \leq -\frac{c}{T-t}\left(\frac{1}{2}\|\xi\|^2 - R^2\right) = \frac{c R^2}{T-t} - \frac{c \|\xi\|^2}{2(T-t)}.
	\end{equation}
	
	Given the bounds $\delta \leq T-t \leq T$, we can further relax the exponential terms to remove the dependency on $t$:
	\begin{align} \label{eq:exp_relaxation}
		\exp\left( \frac{c R^2}{T-t} \right) &\leq \exp\left( \frac{c R^2}{\delta} \right), \\
		\exp\left( - \frac{c \|\xi\|^2}{2(T-t)} \right) &\leq \exp\left( - \frac{c \|\xi\|^2}{2T} \right).
	\end{align}

	Substituting these back into the derivative bound, and applying the polynomial growth of $\rho(\xi)$ from Assumption \ref{assump:constraint}, we obtain a uniform bound for the integrand's derivative on the compact domain $[0, T-\delta] \times \overline{B_R(0)}$:
	\begin{equation} \label{eq:dominating_function}
		|\mathcal{D} f(t,x,\xi)| \leq \frac{K_{\mathcal{D}} K_\rho (1+\|\xi\|^q)}{\delta^{\frac{n+r_{\mathcal D}}{2}}} \exp\left( \frac{c R^2}{\delta} - \frac{c \|\xi\|^2}{2T} \right).
	\end{equation}
	
	By the same argument used in the proof of Lemma \ref{lemma:strict_positivity}, the right-hand side of \eqref{eq:dominating_function} is bounded by $C'_{\delta,R}\exp(-\alpha\|\xi\|^2)$ for some constants $C'_{\delta,R},\alpha>0$. These constants can be chosen uniformly over the finitely many differential operators $\mathcal D$. By Assumption \ref{assump:exp_integrability}, this Gaussian bound is integrable over $\mathcal M_b$. The dominated convergence theorem justifies differentiation under the integral sign. Moreover, since $p_{t,T}$ and its required derivatives are continuous in $(t,x)$ and locally dominated by the same integrable function, applying the dominated convergence theorem again shows that $h^*$ and the resulting derivatives are continuous. Hence, $h^*\in C^{1,2}([0,T)\times\mathbb R^n)$.

\section{Proof of Lemma~\ref{lemma:martingale}}\label{app:martingale}

	For every $t<T$, the Markov property and the Chapman--Kolmogorov identity give
	\begin{align}
		\mathbb{E}^{\mathbb{P}}[M_t]
		&=
		\int_{\mathbb{R}^n}
		h^*(t,w) p_{0,t}(w | x_0) \diff w \notag \\
		&=
		h^*(0,x_0)<\infty,
	\end{align}
	where the last inequality follows from Lemma \ref{lemma:strict_positivity}. Thus, $M_t$ is integrable.

	For $0\leq s<t<T$, the Markov property and the same composition identity yield
	\begin{align}
		\mathbb{E}^{\mathbb{P}}[M_t | \mathcal{F}_s]
		&=
		\int_{\mathbb{R}^n}
		h^*(t,w) p_{s,t}(w | X_s) \diff w \notag \\
		&=
		h^*(s,X_s)=M_s.
	\end{align}
	Hence, $M$ is a martingale. Its strict positivity follows from Lemma \ref{lemma:strict_positivity}, and its path continuity follows from Lemma \ref{lemma:smoothness} and the continuity of $X$.

\bibliographystyle{IEEEtran}
\bibliography{mybibfile}

\ifCLASSOPTIONcaptionsoff
  \newpage
\fi

\end{document}